\documentclass[
 reprint,
superscriptaddress,
floatfix,
 amsmath,amssymb,
 aps,
]{revtex4-1}

\usepackage[utf8]{inputenc}
\usepackage{amsmath,esint}
\usepackage{mathtools}
\usepackage{hyperref}
\hypersetup{colorlinks=true,linkcolor=blue,citecolor=blue,urlcolor=blue}
\usepackage{amsfonts}
\usepackage{amssymb}
\usepackage{bm}
\usepackage{textcomp}
\usepackage{empheq}
\usepackage{siunitx}
\usepackage[titletoc,title]{appendix}
\usepackage{graphicx}
\graphicspath{{all_versions_figure/}}
\usepackage{dcolumn}
\usepackage{bm}
\usepackage{subcaption}
\usepackage{xcolor,soul}
\usepackage{ upgreek }

\newcommand{\citeasnoun}[1]{Ref.~\cite{#1}}

\newcommand{\Figref}[1]{Figure~\ref{fig:#1}}
\newcommand{\figref}[1]{Fig.~\ref{fig:#1}}
\renewcommand{\eqref}[1]{Eq.~(\ref{eq:#1})}

\newcommand{\secref}[1]{Sec.~\ref{sec:#1}}

\newcommand*{\EF}{E_{\rm F}}
\newcommand*{\SiO}{\rm{SiO_2}}
\newcommand*{\epsb}{\epsilon_{\rm{b}}}

\newcommand*{\epsn}{\epsilon_{\rm{0}}}
\newcommand*{\epss}{\epsilon_{\rm{s}}}

\renewcommand*{\wp}{\omega_{\rm{p}}}

\renewcommand*{\wr}{\omega_{\rm{r}}}
\newcommand*{\wn}{\omega_{\rm{0}}}
\newcommand*{\wwien}{\omega_{\rm{Wien}}}
\newcommand*{\meff}{m_{\rm{eff}}}

\newcommand*{\ttheta}{\Theta(\omega,T)}
\newcommand*{\dtheta}{\frac{{\rm d}\Theta(\omega,T)}{{\rm d}T}}

\newcommand*{\kb}{k_{\rm{B}}}

\newcommand*{\knz}{k_{\rm{0,z}}}

\newcommand*{\wAi}{\omega_{{\rm a},i}}
\newcommand*{\wBi}{\omega_{{\rm b},i}}
\newcommand*{\wi}{\omega_{0,i}}
\newcommand*{\sigtwod}{\sigma_{\rm{2D}}}
\renewcommand{\Re}{\operatorname{Re}}
\renewcommand{\Im}{\operatorname{Im}}

\newcommand*{\SM}{SM}

\begin{document}

\title{Optimal materials for maximum near-field radiative heat transfer}

\author{Lang Zhang}
\affiliation{Department of Applied Physics and Energy Sciences Institute, Yale University, New Haven, Connecticut 06511, USA}
\author{Owen D. Miller}
\affiliation{Department of Applied Physics and Energy Sciences Institute, Yale University, New Haven, Connecticut 06511, USA}

\date{\today}

\begin{abstract}
    We consider the space of all causal bulk materials, 2D materials, and metamaterials for maximum near-field radiative heat transfer (RHT). Causality constrains the bandwidth over which plasmonic response can occur, explaining two key traits in ideal materials: small background permittivities (minimal high-energy transitions in 2D materials), and Drude-like free-carrier response, which together optimally yield 10X enhancements beyond the theoretical state-of-the-art. We identify transparent conducting oxides, III-Nitrides, and graphene as materials that should offer nearly ideal near-field RHT rates, if doped to exhibit plasmonic resonances at what we term ``near-field Wien frequencies.'' Deep-subwavelength patterning can provide marginal further gains, at the expense of extremely small feature sizes. Optimal materials have moderate loss rates and plasmonic response at \SI{19}{\mu m} for \SI{300}{K} temperature, suggesting a new opportunity for plasmonics at mid- to far-infrared wavelengths, with low carrier concentrations and no requirement to minimize loss.
\end{abstract}

\maketitle

\section{Introduction}
In this Article, we identify optimal materials and metamaterials for maximum near-field radiative heat transfer (RHT) between large-area planar bodies. We optimize over the space of all causality-allowed material-permittivity or conductivity distributions and discover the possibility for heat-transfer coefficients at the level of \SI{2e5}{W/(m^2K)} at \SI{10}{nm} separations (and \SI{300}{K} temperature), more than 10X higher than the current theoretical state-of-the-art~\cite{enhancement_sio2,Joulain_sio2_2002}. These bounds enable identification of three key characteristics of optimal materials: small background permittivities (or, for 2D materials, minimal high-energy electronic transitions), moderate loss rates, and single-pole Drude-like response with $\approx \SI{19}{\mu m}$ effective surface-plasmon wavelength. These three criteria are not all satisfied by any of the typical bulk materials proposed for near-field RHT; for example, doped Silicon~\cite{IR_Si_2009,dope_Si_nanostructures_2014,si_metasurface_2017} has a large background permittivity, while polar dielectrics~\cite{SPhP_materials_review,nfrht_sphp} have non-Drude-like, highly dispersive narrow-band response. Among bulk materials, we identify transparent conducting oxides (TCOs) and III-Nitrides at low to medium carrier concentrations ($\approx \SI{e18}{cm^{-3}}$) as particularly promising material classes, with the capability to exhibit record RHT rates and to approach within a factor of 2 of the causality-based bounds. Among metamaterials, we show that hyperbolic effective-medium response is nonideal, and that although patterned-cylindrical-hole structures can enable slight enhancements to RHT response, they may require unrealistic feature sizes to do so. We use a gap-surface-mode analysis to provide physical intuition supporting the ideal material characteristics, and we derive a ``near-field Wien's Law'' to prescribe the optimal resonance frequencies at any temperature. Because near-field local densities of states cannot scale with the square of frequency, $\sim\omega^2$, like far-field plane-wave states do, the optimal near-field resonance frequencies are significantly red-shifted relative to the classical Wien frequencies, yielding optimal HTC rates of $(\SI{760}{W/m^2 K^2})T$ as a function of temperature $T$. Interestingly, the optimal causal 2D materials can have slightly superior HTCs to their bulk counterparts, and realistic 2D plasmonic materials at low carrier concentrations or Fermi levels and moderate loss rates also offer the prospect for record-level near-field RHT rates. From a materials perspective, these results offer a new opportunity for plasmonics: instead of pushing for near-zero loss and the highest possible carrier concentrations to exhibit near-visible-frequency resonances~\cite{low_loss_plasmonics}, optimal materials have moderate loss rates and support mid- to far-infrared resonances arising from low to moderate carrier concentrations. More broadly, these optimal characteristics we present can provide guidelines for material choices and designs for a wide range of thermal applications in the near-field, such as thermophotovoltaics~\cite{thinTPVcell2017,nanogap_TPV,InSb_TPV,broadening_emission,TPV_waste_heat}, heat-assisted magnetic recording~\cite{hamr2009,nf_transducer}, nanolithography~\cite{nanolithography_1999}, and thermal management~\cite{thermal_transistor,thermal_rectification,nf_refrigeration,thermal_diode}.

In recent years, near-field RHT rates significantly higher than the blackbody limit have been measured between $\SiO$, SiC, gold, and doped Si in pioneering experiments~\cite{enhancement_sio2,gang_2009,lipson_mems_sic,conductance_dielectric_metal,nfrht_device_si,nanogap_TPV,enhancement_parallel,crossover_conduction_radiation}, inspiring a search for the best material and structure combinations for near-field RHT~\cite{review_nfrht,Miller2015,multilayer_optimization_nfrht,Abdallah_hyperbolic,Abdallah2010limit,limited_role_structuring,dissimilar_nfrht}. Polar dielectrics seem sensible for their strong surface phonon polaritonic resonances in the infrared. $\SiO$ plates, in particular, could in theory yield \SI{300}{K} HTC of about \SI{2e4}{W/(m^2K)} at \SI{10}{nm} separations ~\cite{enhancement_sio2}. Yet, as we will show, the limited bandwidth available in polar dielectrics hinders their ability for further increases in HTC. Traditional plasmonic materials and doped semiconductors have also been explored~\cite{conductance_dielectric_metal,lowT_tungsten,nfrht_device_si}. But in order to tailor their surface plasmon polariton resonance wavelengths to match the optimal ones, the near-field ``thermal wavelengths'' for polaritonic materials have to be derived first, which is one of the goals of our work. Advanced material-growth and nanofabrication techniques have enabled wavelength- to deep-subwavelength-scale patterning of materials~\cite{lipson_mems_sic,dispersion_metasurface_review}. Both hyperbolic metamaterials and in-plane structured metamaterials have been theoretically shown to offer RHT performances better than those of the bulk~\cite{Abdallah_hyperbolic,si_metasurface_2017,multiple_surface_states_2018,nfrht_metasurface_2015,dope_Si_nanostructures_2014}. As predicted by rigorous-coupled-wave-analysis computations, at \SI{300}{K}, air-hole patterned doped Si at carrier concentration of $n= \SI{e20}{cm^{-3}}$ can offer comparable HTC value to that from $\SiO$ at \SI{10}{nm}, and even better relative values for larger separations until \SI{1000}{nm}~\cite{si_metasurface_2017}.

A fundamental question yet to be answered is which material properties enable \emph{maximal} HTC between two extended structures at any given temperature. Specific instances of this question have been explored theoretically and computationally~\cite{tcmt_nfrht,parametric_optimization,joulain_optimization_2013,nfrht_metasurface_2015,dope_Si_nanostructures_2014}. References~\cite{tcmt_nfrht,Chalabi_2014_cmt} provide modal analyses of gap surface waves, though without intuition about integrated broadband radiative response and optimal material properties. Numerical optimizations for single-pole permittivity lineshapes of materials have been done~\cite{parametric_optimization,joulain_optimization_2013}, but without contextualization in the broader landscape of material possibilities, and without guidance relating the optimal single-pole parameters to temperatures, gap distances, and related system parameters. Moreover, none of these works consider optimality criteria of 2D materials. Still they help us better delineate our questions: what is the best material for near-field RHT out of all causality-allowed materials, including multiple-pole bulk materials and 2D materials? Can we theoretically explain the optimal parameters? How do the optimal parameters vary with temperature? We start by doing numerical optimizations in search of the optimal linear permittivity for bulk materials and 2D conductivity for 2D materials at \SI{300}{K}, with passivity as the only constraint (\secref{optimization}). The results provide not only the largest possible HTC, but also intuitions of optimal material characteristics, which for both bulk and 2D materials entails that small-background-permittivity single-Drude-pole plasmonic materials with moderate loss could provide the ideal lineshape, and that the ideal frequency of peak spectral contribution is about \SI{0.067}{eV}. Through gap surface resonance modal analyses, and rigorous computations of HTC, the temperature-independent optimal material properties can be intuitively explained (\secref{lineshape}). As for the optimal resonance frequency (frequency of peak spectral contribution in the case of 2D materials), which varies with the operating temperature, we apply an analysis similar to that leading to Wien's Law for blackbodies, and derive near-field versions that account for the spectral profiles of LDOS. These near-field laws define the optimal resonance frequencies as a function of temperature for bulk materials, with a linear scaling factor significantly smaller than that of a blackbody (\secref{wien}). Optimal HTC also scales linearly with temperature, different from the cubic dependence in the far-field case. Furthermore, we study the effects of deep-subwavelength patterning under the framework of effective medium theory (EMT), and suggest the optimal schemes and parameters for nanostructuring. Large reductions in carrier concentrations can be provided by in-plane patterning of cylindrical air-holes, yielding enhancement in HTC for originally sub-optimal materials. However, originally optimal materials still provide good, if not better HTC values without nanopatterning, in which sense deep-subwavelength patterning only makes sense if one had to begin with highly sub-optimal materials, and feature sizes well below the gap distance can be fabricated (\secref{patterning}). Similar optimizations over all causality-consistent 2D conductivity for 2D materials present findings which are excitingly similar to those of bulk materials and can be explained likewise. Despite quite different gap surface resonance modal dispersions for bulk and 2D materials in the plane--plane configuration, the optimal 2D material is the direct 2D analog of the optimal bulk material, as it has a single Drude pole, a moderate loss rate, and minimal higher-energy electronic transitions. (\secref{materials2D}). Among common materials, we predict that TCOs, III-Nitrides with small background permittivities, bulk or 2D, as well as other 2D doped semiconductors and 2D semimetals with predominantly single-Drude-pole 2D conductivity, once synthesized and engineered to possess low to medium carrier concentrations and moderate loss levels, could potentially approach the optimal permittivity or the optimal 2D conductivity, and yield HTC 5X better than $\SiO$ for a wide range of gap separations (\secref{realmats}).

\section{Ideal causality-allowed $\epsilon(\omega)$ for maximum HTC} \label{sec:optimization}
In this section, we formulate the optimization of heat-transfer coefficients (HTCs) over all causality-allowed material permittivities $\varepsilon(\omega)$. We start with the standard expressions for computing HTC via modal-photon-exchange functions (\secref{HTCplanar}) and list common permittivity lineshapes (\secref{common}, Drude, Drude--Lorentz, etc.). In \secref{opt}, we describe the Kramers--Kronig-based representation of all causality-consistent permittivities, and we show that numerical optimizations identify key material characteristics that are optimal for near-field RHT. We concentrate on HTC in this section to isolate the effects of a single temperature, which makes it useful for comparisons across the literature, and as discussed in \secref{wien} our results translate seamlessly to RHT between two bodies with arbitrary temperature differences.



\begin{figure*}[htb]
  \includegraphics[width=\textwidth]{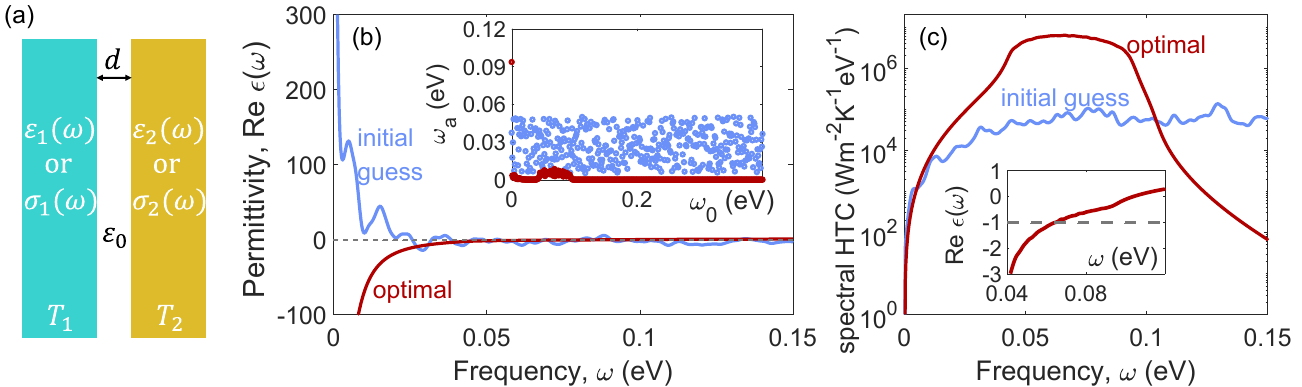}
  \caption{Numerical optimizations of permittivity profiles to maximize HTC of plane--plane configuration at 300K. (a) Schematic of the plane--plane configuration with bulk materials represented by $\epsilon(\omega)$ and 2D materials by $\sigma(\omega)=\sigma_{\rm 2D}(\omega)$. (b) The initial guess (blue) and optimal permittivity profile (red) of one representative 400-oscillator optimization, with a grey dashed line indicating $\epsilon = -1$. Inset shows the corresponding oscillator amplitudes $\wAi$. From a random starting point where all $\wAi\neq 0$, the optimization result has the largest non-zero amplitude for the Drude oscillator and only a few small amplitudes for Drude--Lorentz oscillators at low frequencies. The resultant lineshape is predominantly that of a single-pole Drude permittivity with $\omega_{\rm{a},0}=\wp=\SI{0.094}{eV}$. (c) Spectral HTC from the initial (blue) and optimal (red) permittivities. Inset: zoomed-in profile of the optimal $\epsilon_{\rm r}$ near the near-field thermal frequency at \SI{300}{K}, which is derived in \secref{wien}.}
  \label{fig:optimization}
\end{figure*}

\subsection{HTC between planar structures}
\label{sec:HTCplanar}
The canonical configuration of extended near-field RHT consists of two parallel half-spaces (or two parallel planar structures for 2D materials) separated by a vacuum gap with thickness $d$ much smaller than the characteristic thermally excited wavelength~\cite{planar_nfrht,nanolithography_1999,Joulain_sio2_2002}. For materials with translation and rotation symmetry in the plane parallel to the surfaces (isotropic or anisotropic out of plane), HTC is calculated with a double-integration over all plane-wave channels at frequency $\omega$ and surface-parallel wavenumber $\beta$. The infinitesimal temperature difference between the two bodies manifests in a temperature derivative of the Planck distribution $\ttheta$, where $\ttheta = \hbar\omega / \left(e^{\hbar\omega/\kb T}-1\right)$ and $\kb$ is the Boltzmann constant. Then the HTC is given by~\cite{surface_wave_thermal}
\begin{equation}
    {\rm HTC} = \frac{1}{4\pi^2}\int_0^{\infty}\dtheta \underbrace{\int_0^{\infty}\xi(\omega,\beta)\beta{\rm d}\beta}_{S(\omega)} {\rm d}\omega,
    \label{eq:HTC}
\end{equation}	
for any polarization. For evanescent waves, the variable ${\xi(\omega,\beta)} = \frac{4\Im{r_{01}}\Im{r_{02}}e^{2i\knz d}}{|1-r_{01}r_{02}e^{2i\knz d}|^2}$ is a modal photon exchange rate that is proportional to the transmission probability through each plane-wave channel,
with the $z$-component of wavevector in vacuum denoted as $\knz=\sqrt{\frac{\omega^2}{c^2}-\beta^2}$ and reflectivity from vacuum gap to medium 1 or medium 2 as $r_{01,2}$. We refer to the wavenumber-integrated quantity $S(\omega)$ as the ``spectral photon exchange.''


\subsection{Common permittivity lineshapes}
\label{sec:common}
Metals, doped semiconductors, and polar dielectrics are all materials of interest for large near-field RHT. Their permittivity lineshapes arise from electronic transitions (intraband and interband), optical phonons, and related processes determining optical properties for most materials~\cite{yu_2016,kaxiras_joannopoulos_2019}.

The simplest Drude lineshape can often describe \emph{intra}band transitions, and is given by~\cite{kaxiras_joannopoulos_2019,bound_material_loss}
\begin{equation}
\epsilon(\omega) = \epsb\left(1-\frac{\wp^2}{\omega^2+i\omega\gamma} \right).\\
    \label{eq:Drude}
\end{equation}	
The background permittivity $\epsb$ arises from electronic transitions with frequencies much higher than thermally interesting ones. (Whether it multiplies the second term or not amounts to a simple redefinition of the parameter $\wp$.) The plasma frequency $\wp$, the amplitude or strength of the oscillator, is a measure of free-carrier density $n$ (generalized to incorporate background permittivity and effective mass):
\begin{equation}
\wp^2 = \frac{ne^2}{\epsb\meff},  \\
    \label{eq:wp}
\end{equation}
where $\meff$ is the free-carrier effective mass. From electron scattering rate $\gamma$ one can define a dimensionless loss rate $g = \gamma / \wp$.

A Drude--Lorentz lineshape, which describes, for example, \emph{inter}band transitions and optical-phonon contributions, is given by~\cite{yu_2016,SPhP_materials_review}
\begin{equation}
\epsilon(\omega) = \epsb\left(1-\frac{\omega_{\rm a}^2}{\omega^2-\wn^2+i\omega\gamma} \right), \\
    \label{eq:DL}
\end{equation}	
where the frequency $\omega_0$ is the band-to-band transition frequency (or transverse-optical-phonon frequency) and the oscillator strength is now denoted by $\omega_{\rm a}$. Knowledge of the ratio of the static dielectric constant $\epss$ to $\epsb$ specifies the ratio $\omega_{\rm a}^2 / \omega_0^2$, which follows from \eqref{DL} and is known as the Lyddane-Sachs-Teller relation~\cite{LST_relation}.


\subsection{Numerical optimization of $\epsilon(\omega)$ for HTC at \SI{300}{K}}
\label{sec:opt}
Any causal physical material permittivity must satisfy the Kramers--Kronig relations~\cite{lucarini_2010} that relate the real part $\Re \epsilon$ at one frequency $\omega$ to an integral of the imaginary part $\Im \epsilon$ (or vice versa) over all frequencies ($\omega'$):
\begin{align}
    \Re \epsilon(\omega) = 1 + \frac{2}{\pi} \int_0^{\infty} \frac{\omega' \Im \epsilon(\omega')}{(\omega')^2 - \omega^2} \,{\rm d}\omega',
    \label{eq:KK}
\end{align}
where the integral is a principal-value integral. Notice the suggestive form of the integrand of \eqref{KK}, which is similar to a lossless Drude--Lorentz oscillator with integration variable $\omega'$ as the effective transition frequencies of a continuum of oscillators. As we show in the {\SM}, this correspondence can be formalized: for a discretization of the Kramers--Kronig relation into local basis functions, one can write the permittivity as a sum of Drude--Lorentz oscillators with infinitesimal loss rates:
\begin{equation}
\epsilon(\omega) = 1 - \sum_{i=1}^N \frac{\wAi^2}{\omega^2-\wi^2+i\omega\gamma_i},  \\
    \label{eq:eps_repre}
\end{equation}
where $\gamma_i \rightarrow 0$ from above. Such a representation is completely general, and applies for arbitrarily high loss levels in a material. (Lossless Drude--Lorentz oscillators have delta-function imaginary parts with arbitrarily large amplitudes, which can be derived from \eqref{eps_repre} in the $\gamma_i \rightarrow 0$ limit, cf. {\SM}.) A similar representation can be derived via a Mittag--Leffler expansion, albeit with possibly lossy oscillators and without consideration of higher-order poles~\cite{mittag_optimizing,unified_dispersion}. In some scenarios, one may use alternative oscillator types (e.g. Gauss--Lorentz~\cite{causal_permittivity}) which, for a \emph{small} number of oscillators, may be a better approximation of certain dielectric functions. But ultimately any such permittivity must be representable by \eqref{eps_repre}.

Optimizing HTC over all causality-allowed permittivities can then be done by optimizing \eqref{HTC} over all possible oscillator strengths $\wAi$ and frequencies $\wi$ (with infinitesimal loss rates $\gamma_i$) for the two materials involved. We allowed for the possibility of different materials for the two bodies, but the optimizations always converged on identical permittivity profiles. For a single material, we optimize over these parameters by choosing a large number (hundreds) of oscillator frequencies $\wi$ to cover the full relevant bandwidth. Then we do a gradient-descent-based local optimization of \eqref{HTC} over the corresponding hundred-plus $\wAi$ values, using semi-analytical expressions of the HTC gradients with respect to all parameters (cf. {\SM}). To avoid poor-quality local optima, we typically run the optimizations in two stages: first with a smaller number of parameters starting from a random initial guess, and then using ``successive refinement''~\cite{robust_successive_refinement} to polish the optimal solution with a very large number of parameters (cf. {\SM} for more details). The optimizations typically converge within a few hundred iterations. \Figref{optimization} depicts the initial and final permittivity distributions of a 400-oscillator optimization, with a random initial permittivity profile (blue) that converges to a smooth Drude-dominant profile by the end of the optimization. We repeated this process with many random starting points and found that lineshapes nearly identical to that shown in \figref{optimization} appear to be globally optimal or nearly so. The ideal lineshape has a Drude pole ($\wi=0$) with large oscillator strength, $\wAi=\SI{0.094}{eV}$, that is the dominant feature of the lineshape. (The optimal oscillator strength scales with temperature, as discussed in \secref{wien}.) The inset of \figref{optimization} shows the oscillator weights: there are typically a few other low-energy oscillators ($\wi<\SI{0.1}{eV}$) with small but nonzero oscillator strengths, which provide small adjustments to the lineshape to broaden the resonant bandwidth, with all remaining oscillator strengths converging to zero (cf. {\SM}). The HTC from such simple lineshapes is approximately $\SI{2.6e5}{W/(m^2 K)}$, a record level for \SI{10}{nm} separations and \SI{300}{K} temperatures.

Natural questions, then, are why the Drude lineshape is superior to more complex possibilities, why the optimal oscillator strength is $\SI{0.094}{eV}$, and whether real materials can approach the optimal HTC values. In the next sections, we develop simple theoretical explanations of these questions. We start with the question of the optimal lineshape.

\section{The ideal permittivity profile: Drude with small $\epsb$ and moderate $g$}
\label{sec:lineshape}
In this section we pinpoint the physical underpinnings for the optimal Drude lineshape. We focus on three ideas: why Drude is better than Drude--Lorentz or more complex lineshapes, why a small background permittivity is better than large ones, and why moderate loss rates are also important. For the first two, minimal dispersion in the permittivity is the key controlling factor; it is not possible for a material to exhibit a resonant surface-plasmon permittivity of $\Re \epsilon \approx -1$ over an infinite bandwidth (also due to causality constraints~\cite{Gustafsson_2010}), but Drude materials with small background permittivities appear to offer the largest possible such bandwidths. With respect to the loss rate, moderate losses are optimal because there are tradeoffs between the source amplitudes and resonant amplifications that result in specific, moderate ranges of optimal loss rates.
\begin{figure}[htb]
  \includegraphics[width=0.5\textwidth]{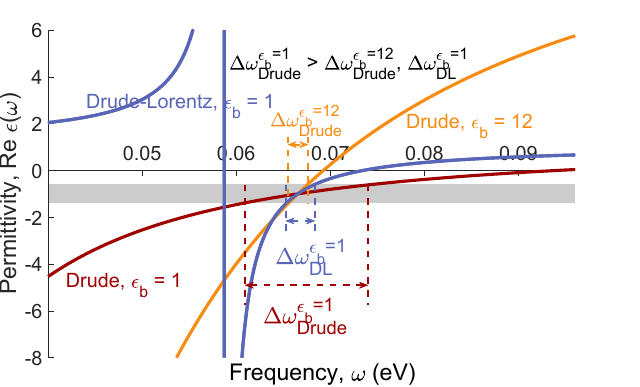}
  \caption{Three representative classes of permittivity lineshapes. Red: Drude with background permittivity $\epsb=1$, orange: Drude with $\epsb=12$, blue: Drude--Lorentz with $\epsb=12$, all with equal loss rate $g = 0.1$. Near-resonance frequencies with $\Re \epsilon \in [-0.6,1.4]$ are shaded in grey to indicate the high-rate bandwidths of each lineshape. Drude with $\epsb=1$ provides much broader bandwidth with $\Re \epsilon \approx -1$ than the other two lineshapes.}
  \label{fig:eps_profile}
\end{figure}

\subsection{Small background permittivities}
\label{sec:bgepsilon}
\begin{figure*}[htb]
  \includegraphics[width=\textwidth]{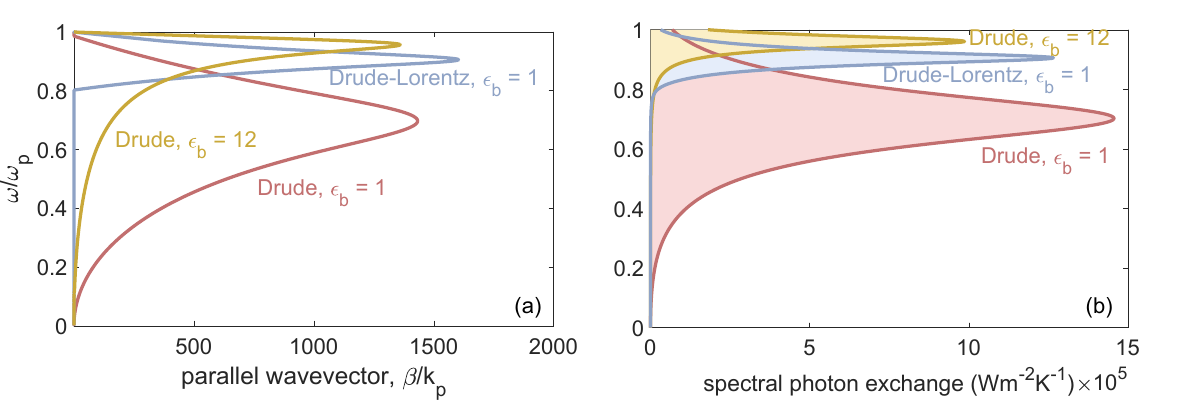}
  \caption{(a) Modal dispersions and (b) spectral photon exchanges of three representative lineshapes at their optimal loss-rates: Drude with $\epsb=1$ (red), Drude with $\epsb=12$ (yellow) and Drude--Lorentz with $\epsb=1$ (blue), with frequencies normalized to their respective oscillator frequencies. Both panels show that Drude with $\epsb=1$ lineshape gives broadest-band LDOS and therefore heat transfer profiles.}
  \label{fig:types}
\end{figure*}
\begin{figure*}[htb]
  \includegraphics[width=\textwidth]{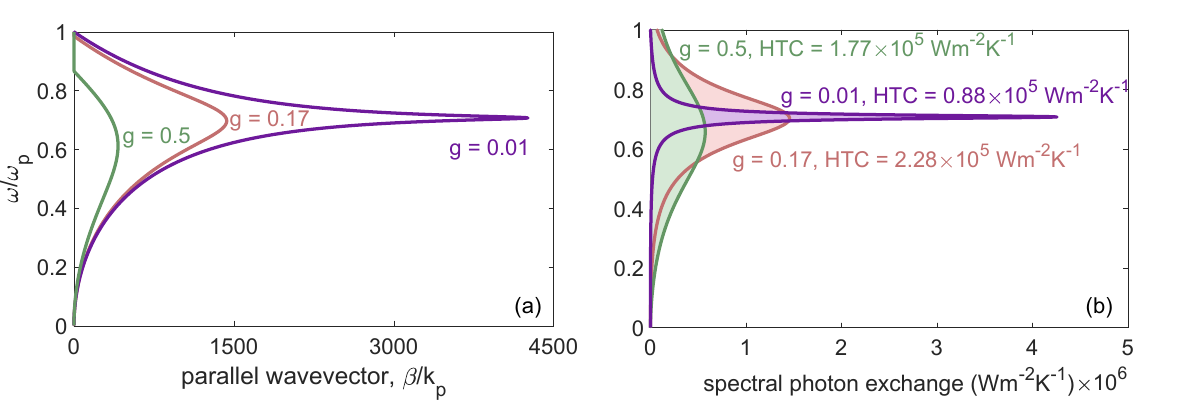}
  \caption{Spectral dispersions and photon exchanges of Drude with $\epsb=1$ at three different loss rates, with frequencies normalized to plasma frequency $\wp$. (a) Modal dispersions of gap surface waves. The permittivity with loss rate too large (green) suffers from poor spatial confinement. Smaller loss rates (purple) correspond to larger in-plane wavevector and better confinement on resonance, yet according to (b) the spectral photon exchanges, too small of a loss rate results in overly narrow spectral bandwidths. A moderate loss rate (red) balances the effects of spatial confinement and spectral bandwidth and gives best total HTC from plasmonic materials.}
  \label{fig:gamma}
\end{figure*}
The surface wave at an interface between a polaritonic material and air exhibits a modal dispersion relation for the wavenumber $\beta$ given by~\cite{maier_2007}
\begin{equation}
    \beta = \frac{\omega}{c}\sqrt{\frac{\epsilon(\omega)}{1+\epsilon(\omega)}}. \\
    \label{eq:beta}
\end{equation}
The largest confinement, which occurs for the largest $\beta$, occurs as $\epsilon(\omega)$ approaches -1 from below, in the low-loss limit. This condition also holds for two-interface geometries, e.g. metal--insulator--metal, where in the high-wavenumber limit the two interfaces effectively decouple~\cite{maier_2007}. Moreover, one can show (cf. {\SM} of \citeasnoun{Miller2015}) that the same condition of $\Re \epsilon(\omega) \approx -1$ is the condition at which peak HTC occurs between planar layers in the low-loss limit, as the high confinement leads to the strongest resonant energy transfer. Thus maximum HTC requires the largest possible bandwidth over which $\Re \epsilon(\omega) \approx -1$. (Or, more precisely~\cite{Shim2020}, for which $\Re (-1/(\epsilon-1)) \approx -1/2$.) This bandwidth cannot be infinite: causality, again manifest through the Kramers--Kronig relations, dictates limits to the largest bandwidth for which a specific negative permittivity can be achieved~\cite{Gustafsson_2010}. Our numerical computations of \secref{opt} imply that Drude lineshapes offer nearly the largest possible bandwidth for which $\Re \epsilon(\omega) \approx -1$.

\Figref{eps_profile} demonstrates why a Drude material with small background permittivity provide the largest bandwidth with $\Re \epsilon(\omega) \approx -1$. Three permittivity lineshapes are depicted: Drude with small background permittivity (red), Drude with large background permittivity (orange), and Drude--Lorentz with small background permittivity (blue). The shaded grey region covers real permittivity values between $-1.4$ and $-0.6$, for clear visual indication of bandwidth. The Drude--Lorentz bandwidth is quite small due to the nonzero transition frequency; $\Re \epsilon$ ascends from $-\infty$ at a nonzero frequency, and does so much faster than a Drude material, exhibiting large dispersion and thus small bandwidth. A large background permittivity has a similar effect. As can be seen in \figref{eps_profile}, as well as from \eqref{Drude}, a large background permittivity increases the slope of the permittivity lineshape at every frequency, hence increasing its dispersion and reducing its bandwidth. By contrast, the Drude lineshape with small background permittivity exhibits the least amount of dispersion and the largest bandwidth.

Further quantitative support of the importance of Drude-type response and small background permittivity is given in \figref{types}. \Figref{types}(a) shows the modal dispersion relations~\cite{maier_2007} between two half spaces with the three lineshapes as in \figref{eps_profile}, each with its optimal loss rate in terms of HTC. \Figref{types}(b) shows the spectral photon exchange, i.e. the temperature-independent part of the HTC integrand. The close correspondence between (a) and (b) confirms the suitability of using modal analysis to interpret the modal and spectral components of HTC. At their respective optimal loss rates, the largest achievable $\beta$ values are similar, indicating similar levels of spatial confinement on resonance. Yet one can see that the larger bandwidth of the small-background-permittivity Drude material provides a substantial advantage over the other materials.

Large, nonideal background permittivities usually occur in heavy elements or their compounds, which have many high-energy inner-shell electrons. This indicates the possible superiority of \emph{light} materials, with small attendant background permittivities, a suggestion that is substantiated in our investigation of optimal real materials in \secref{realmats}.



\subsection{Moderate loss}
\label{sec:loss}
The second key factor of the optimal lineshape is a moderate loss rate $\gamma$. \Figref{gamma} compares modal dispersions and spectral photon exchanges among three different choices of $g = \gamma/\omega_{\rm p}$: 0.01, 0.17, and 0.5. Large $g$ broadens the spectral contributions, but gives few high-$\beta$ states or small LDOS on resonance, seen from the dispersion. Meanwhile, small $g$ provides huge LDOS on resonance, at the cost of very narrow peak-HTC bandwidth. The very large loss rates, while penalized less than very small loss rates, show sub-optimal HTC due to their small peak values. The best integrated broadband response comes from intermediate values of $g$, which equals 0.17 for gap separation of \SI{10}{nm}. Our result of moderate loss as optimal confirms coupled-mode predictions of such a trend~\cite{multiple_surface_states_2018}.

\begin{figure*}[htb]
  \includegraphics[width=\textwidth]{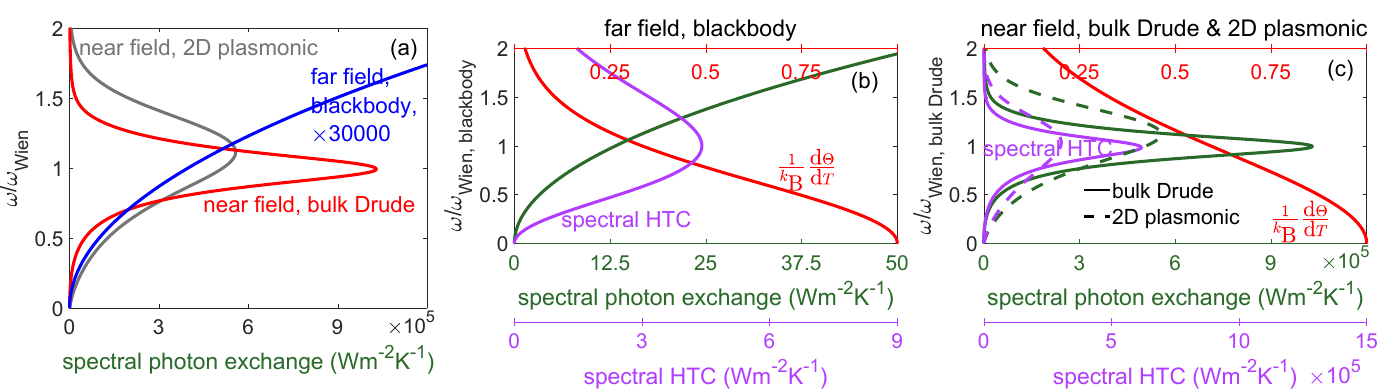}
  \caption{Spectral photon exchanges and spectral HTCs leading to Wien frequencies between different material models: blackbodies, bulk Drude materials and 2D plasmonic materials, normalized to their respective $\wwien$. (a) Comparisons of spectral photon exchanges. In the case of blackbodies (blue), it's directly proportional to the DOS of propagating waves, whereas for bulk (red) and 2D (grey) material, photon exchange profiles are much narrower, and are closely related to LDOS in the near-field. Wien frequency analysis for blackbodies (b), bulk Drude and 2D materials (c). The red curves show the normalized temperature factor $\frac{1}{\kb}\dtheta$, the green curves are the spectral photon exchanges, and the purple curves represent the product of the them, also defined as the spectral HTC. The peaks of the spectral HTC define $\wwien$ for each case.}
  \label{fig:overlap}
\end{figure*}

\section{Wien frequencies for near-field energy exchange between real materials} \label{sec:wien}
The final optimal Drude parameter to explain is the plasma frequency $\wp$, whose optimum is \SI{0.094}{eV} for 300K HTC. This optimal value is closely linked to the precise temperature spectrum of the sources under consideration, and in this section we derive a linear relation between the two.

For two macroscopic bodies in the far field, Wien's displacement law states that the radiated energy of a blackbody is maximized at a frequency linearly proportional to its temperature~\cite{mehra_rechenberg_2001,wien_peak_wavelength}. The radiated energy is $H = (1/4\pi^2) \int_0^\infty \Theta(\omega,T) S(\omega)\,{\rm d}\omega$, where $S(\omega)$ in this case is interpreted as a photon emissivity rate. A key distinction in the far-field case is that the term $S(\omega)$ is proportional to the photon density of states, which scales as the square of frequency, $\sim \omega^2$~\citeasnoun{christensen_2019}). This quadratic scaling is critical to the determination of the Wien peak, blue-shifting the maximum-emission peak relative to the peak of the Planck distribution. One can similarly define a ``Wien'' peak for maximum far-field HTC, simply replacing $\ttheta$ with $\dtheta$. The optimal frequency for maximum HTC again scales linearly with temperature. One can generalize this further to radiative heat transfer between two bodies, with temperatures $T_1$ and $T_2$, yielding slight corrections to the linear relationship. From Wien's Law, the intuition has developed that the thermal wavelength is about 8--\SI{10}{\um} near \SI{300}{K} temperatures ~\cite{wien_peak_wavelength}, but this intuition is only valid for the far field. In the near field, the wavelengths for peak thermal exchange are significantly longer.

In the near field, the spectral photon exchange (which replaces blackbody emissivity in the frequency integral) can exhibit extraordinarily large peaks due to the access to high-confinement near-field waves, but it cannot exhibit scaling $\sim \omega^2$. In contrast to the divergent density of propagating photon states, there is a known sum rule requiring that the integrated near-field \emph{local} density of states must be finite~\cite{Shim2019bandwidth,Manjavacas_LDOS}. For a Drude material, the spectral photon exchange and the local density of states exhibit peaks at a resonant frequency $\wr$ where the real part of the permittivity is -1, i.e. $\Re \epsilon(\wr) = -1$, as seen in the red curve of \figref{overlap}(a). For room temperature and higher, the bandwidth of the Drude-material photon transmission is typically much smaller than the width of the thermal spectrum, such that the overlap between the two is essentially the integral of the spectral photon exchange multiplied by the value of the Planck distribution at $\wr$, contrasted in \figref{overlap}(b,c). This implies that the optimal $\wr$ will maximize the product of two quantities: the integrated energy exchange, and the value of $\Theta(\wr)$. The Planck distribution actually peaks at zero frequency. Yet a Drude material with infinitesimal resonant frequency will necessarily have near-zero bandwidth, and is nonideal. At higher frequencies, the bandwidth increases, though the Planck distribution starts to decrease. In the SM, we show that one can derive transcendental equations relating the optimal $\wr$ for heat transfer, which defines the Wien frequency $\omega_{\rm W}$, relative to temperature, leading to an HTC ``near-field Wien frequency,'' $\omega_{\rm W}^{\rm HTC}$, given by
\begin{align}
    \frac{\hbar \omega_{\rm W}^{\rm HTC}}{k_B T} = 2.57\,.
    \label{eq:hWien}
\end{align}
This optimal resonance frequency directly determines the optimal plasma frequency: $\omega_{\rm p} = \omega_{\rm W}^{\rm HTC} \sqrt{(\epsb+1)/\epsb}$; if we insert \eqref{hWien} into this relation and choose $\epsb = 1$, we find an optimal plasma frequency of \SI{0.094}{eV}, exactly matching that discovered by the computational optimization of \secref{opt}. Moreover, with these optimal resonance frequencies and the optimal lineshape, the maximum possible HTC values scale linearly with temperature, given by
\begin{align}
    {\rm HTC} = \left(\SI{760}{W/m^2/K^2}\right) T.
\end{align}
Thus we have an explanation for the optimal plasma frequency and for the maximal HTC value, which is determined by the combination of bringing a large integrated spectral photon exchange as close as possible to zero frequency, where the Planck distribution peaks.

\begin{figure*}[htb]
  \includegraphics[width=\textwidth]{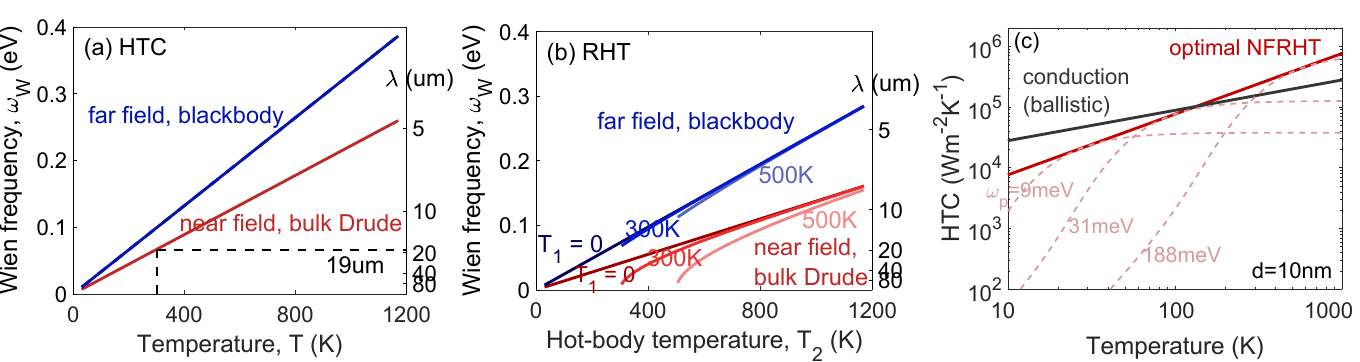}
  \caption{Temperature scaling laws for Wien frequencies for blackbodies in the far field (blue) and bulk Drude materials in the near field (red), and for near-field radiative HTC and conductive HTC. (a) Wien frequencies for HTC at different operating temperatures. For bulk Drude materials in the near-field, $\wwien$ corresponds to \SI{19}{\mu m} at \SI{300}{K}. (b) Wien frequencies for RHT at different hot-body temperatures with selected cold-body temperatures at $T = 0,\ 300,\ 500 \rm{K}$. Blackbody Wien frequencies is always quite larger than those for near-field Drude materials in each of the cases. (c) Comparison of temperature scaling of HTCs of optimal near-field RHT and conductive heat transfer in the ballistic regime. Each dashed line is an HTC curve from an actual material optimized for temperature $T=30, \ 100, \ 600 \ \rm{K}$.}
  \label{fig:wien}
\end{figure*}

At high enough temperatures, for example $T\gtrsim \SI{130}{K}$ for the $d=\SI{10}{nm}$ configurations, these maximum near-field radiative HTCs exceed conductive HTCs in the ballistic regime under standard conditions for temperature and pressure~\cite{many_body_rht}(cf. {\SM}), as shown \figref{wien}(c). Any given material (dashed lines) will not show linear temperature scaling itself, but the envelope of optimal materials (solid red line) exhibit exactly the predicted linear scaling.

A similar analysis can be done for radiative heat transfer (RHT) between two bodies of the same material at temperatures $T_1$ and $T_2$, with $T_2 > T_1$. For $T_1 = 0$ (an exact analog of the conventional Wien-law condition), the optimal Wien frequency $\wr$ for near-field RHT is given by (cf. {\SM})
\begin{align}
    \frac{\hbar \omega_{\rm W}^{\rm RHT}}{k_B T} = 1.59.
    \label{eq:hWienRHT}
\end{align}

In the near field, the optimal Wien frequency is equivalent to the optimal Wien wavelength, regardless of whether the integrand is written in terms of frequency or wavelength. This stands in stark contrast to the far-field case, where the Wien peak is different in the two cases, due to the inverse relationship between the two that enters the differential in addition to the integrand itself~\cite{wien_peak_wavelength}. This does not occur in the near field thanks to the analytical structure of the HTC and heat-transfer expressions in which the integration parameter is effectively $\ln \omega$; since ${\rm d}(\ln \omega) = |{\rm d}(\ln\lambda)|$, there is no distinction whether parametrizing the radiation laws with $\omega$ or $\lambda$.

\Figref{wien} plots the optimal Wien frequencies, and their corresponding Wien wavelengths, for both near- and far-field HTC (a) and RHT (b). One can see the linear scaling relations that emerge in the near-field cases, and the significantly smaller slopes that lead to much longer optimal wavelengths than for far-field blackbodies. Even for temperatures as high as \SI{1000}{K}, the optimal resonance wavelength is $\SI{6}{\mu m}$. For HTC at \SI{300}{K}, the optimal plasma frequency of \SI{0.094}{eV} translates to an optimal Wien frequency of \SI{0.067}{eV}, which corresponds to an optimal operating wavelength of $\SI{19}{\mu m}$.

\section{Deep-subwavelength, possibly anisotropic, structured metamaterials}
\label{sec:patterning}
Metamaterials, which exhibit effective properties different from their constituent materials~\cite{smith_metamaterials}, are a natural platform for potentially achieving maximal near-field HTC and RHT. Metamaterials with isotropic effective permittivities naturally fall under the umbrella of \eqref{eps_repre} and may exhibit HTC values close to the optimal \SI{2e5}{W/m^2 K} at \SI{10}{nm} separations and \SI{300}{K} temperatures, but cannot surpass them. Yet anisotropic effective permittivities, as seen e.g. in hyperbolic metamaterials~\cite{Abdallah_hyperbolic,Zubin_hyperbolic_broadband,dope_Si_nanostructures_2014}, are not subject to the isotropic representation of \eqref{eps_repre} and could potentially exhibit superior performance.

We performed an extensive set of computational optimizations of three classes of Maxwell--Garnett effective-medium metamaterials, cf. {\SM}, and summarize our findings according to the class of structures under consideration:
\begin{enumerate}
\item Isotropic, periodic holes: the materials retain isotropic optical properties. With air fill fraction represented by $f$, the effective permittivity is:
\begin{equation}
\epsilon_{\rm{eff}} = \frac{(1-f)\epsilon+1+f}{1-f+(1+f)\epsilon}\epsilon. \\
    \label{eq:nanohole}
\end{equation}	
As detailed in the supplementary, reducing the electronic density of plasmonic materials brings down both the effective $\epsb$ and effective $\wp$. For non-ideal materials with $\epsb$ and $\wp$ that are too high, isotropic air holes can greatly help. They cannot improve materials that are already ideal.
\item Periodic cylinders: cylinders oriented perpendicular to the surfaces lead to a difference in the permittivities along the ordinary and  extraordinary axes. The ordinary-axis permittivity is the same as that for isotropic periodic holes, and many resonance properties are similar. There can be a little improvement from  anisotropic bands, but the effect is minor and not obvious.
\item Thin-film stacks (hyperbolic metamaterials): deep-subwavelength, periodic multi-layer stacks lead to hyperbolic dispersion bands. However, the hyperbolic resonances are far less tightly confined to the surface compared to plasmon or phonon surface polaritons, resulting in smaller HTC. The thin-film stacks are inferior to periodic cylinders.
\end{enumerate}



\begin{figure*}[htb]
  \includegraphics[width=\textwidth]{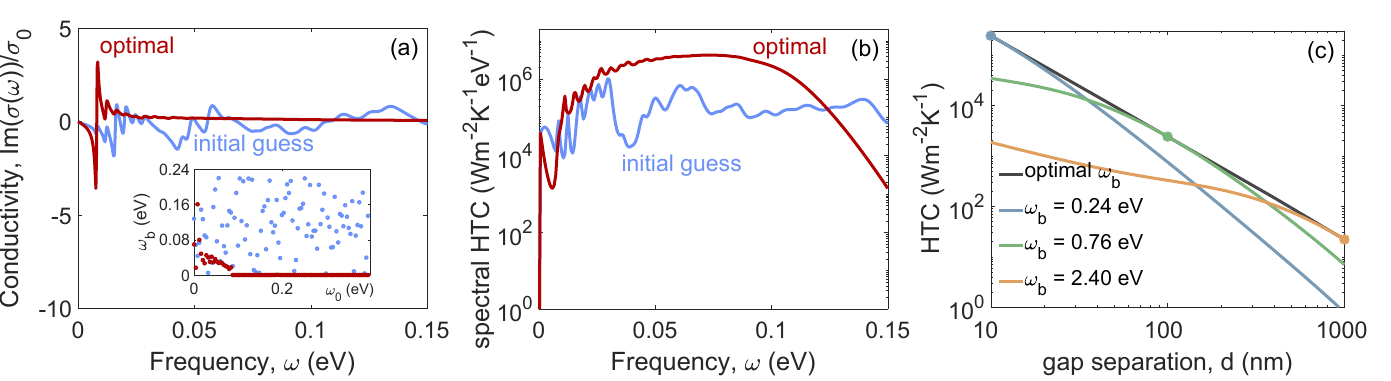}
  \caption{Optimal 2D materials for near-field RHT. (a) The initial guess (blue) and optimal imaginary part of the conductivity profile (red) of one representative optimization, normalized by $\sigma_0 = \frac{e^2}{\pi\hbar}$. Inset shows the corresponding oscillator amplitudes $\wBi$. From a random starting point where all $\wBi\neq 0$, the optimization result has the largest non-zero amplitude for a Drude--Lorentz oscillator with a very small intrinsic frequency ($\approx 10 meV$) and a few small amplitudes for other oscillators at low frequencies. The resultant lineshape is predominantly that of a single-pole Drude--Lorentz conductivity. (b) Spectral HTC from the initial (blue) and optimal (red) conductivities. The optimal profile gives larger spectral contribution over a broad range of resonant frequencies. (c)  Optimal single-pole 2D materials are nearly as good as the optimal multi-pole materials. The optimal Drude-oscillator amplitudes vary with gap separation, in contrast to bulk Drude materials. Optimal parameters for $d=10, \ 100, \ 1000 \ \rm{nm}$ give the gap separation dependence of HTC as the blue, green and yellow curve. Optimal HTC from the optimal $\omega_{\rm b}$ at every $d$ is presented as the black line.}
  \label{fig:optimization_2D}
\end{figure*}

We refer to the {\SM} for a detailed effective-medium theory (EMT) description of the effective permittivities arising from these patterning schemes and a comparison of their optimal HTC values. In \secref{realmats} we compare estimates of optimal nanostructured materials to optimal bulk materials. Our central finding is that patterning can provide marginal improvements, but at the expense of significant fabrication complexity due to the tiny required feature sizes. Moreover, all of the effective-medium-theory values become approximate (and likely overestimates) at the small separation distances of primary interest, further diminishing the possible improvements via such patterning.

\section{2D Materials}
\label{sec:materials2D}
Reduced material dimensionality leads to qualitatively different polaritonic response. The plasmon-polariton dispersion relation for a 2D plasmonic material~\cite{graphene_RHT}, for example, is quite different from that of a bulk plasmonic half space~\cite{maier_2007}, with the resonant frequency scaling as the square root of the wavenumber~\cite{christensen_2019}, $\wr \sim \sqrt{\beta}$, instead of asymptotically approaching a constant value. Yet we still find that optimal 2D materials exhibit narrow-band RHT response relative to blackbodies, offering many similarities to their optimal bulk counterparts. In this section, we carry out similar numerical optimizations and analyses as those in the bulk case, and comparatively present the results for 2D materials by identifying the direct analogs and highlighting the differences with the optimal bulk Drude materials.

\begin{figure*}[htb]
  \includegraphics[width=0.8\textwidth]{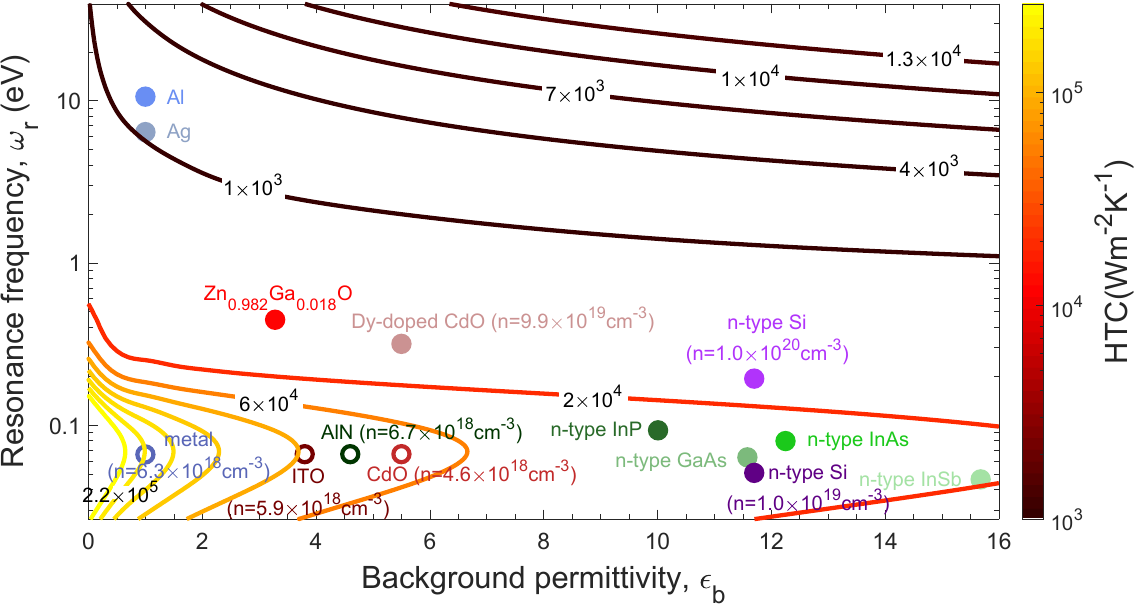}
  \caption{Background permittivity $\epsb$ and resonance frequency $\wr$ for ideal, hypothetical bulk materials (hollow circles) and materials reported in literature (solid circles), on top of which HTC contours with levels ranging from \SI{1e3}{W/(m^2K)} to \SI{2.4e5}{W/(m^2K)}. Permittivity parameters including $\epsb$ and carrier concentrations $n$ are from references~\cite{LD_BB_metal,AZO_GZO_extrapolate,AlN_phonon,IR_Si_2009,si_doping,III-V_FIR,inas_doping_wp_law_2013,insb_bandedge_wp_law_2014,cdody_2015}.}
  \label{fig:epsb_wr}
\end{figure*}

\begin{figure*}[htb]
    \includegraphics[width=0.85\textwidth]{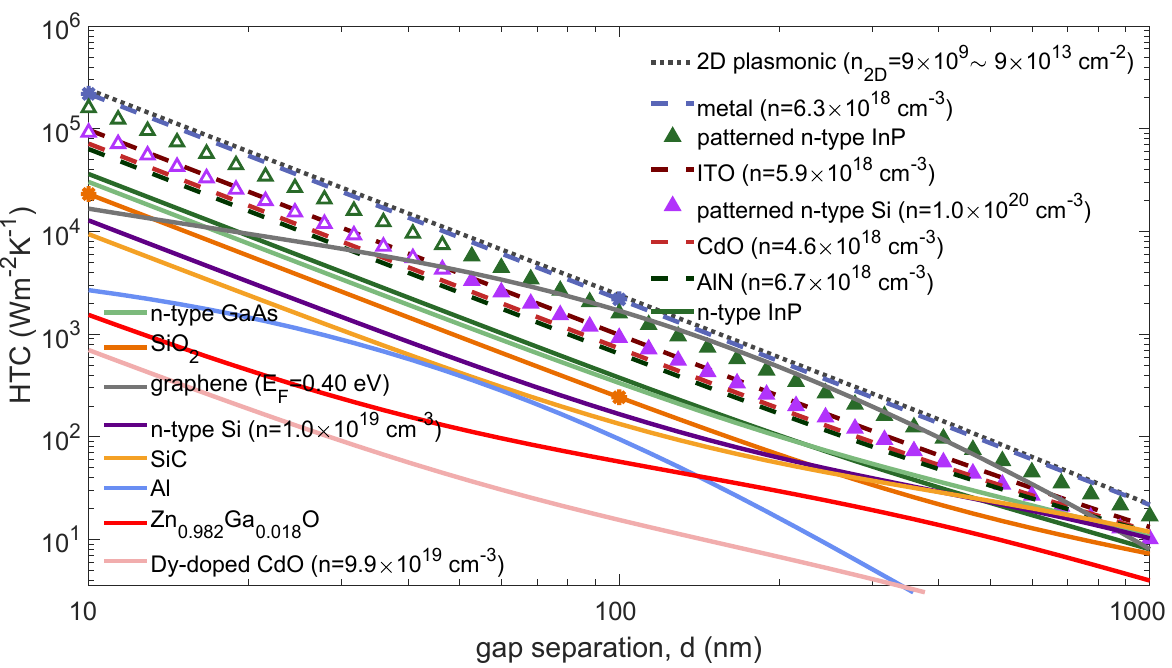}
  \centering
  \caption{HTC values at \SI{300}{K} of the plane--plane configuration at different gap separations $d$ for ideal bulk materials (dashed lines) and 2D materials (dotted lines), materials with experimentally measured permittivity or conductivity data available (solid lines), and materials with periodic-cylindrical-hole patterning (triangles). The two orange star markers on top of the $\SiO$ line mark the previous theoretical state-of-the-art of unpatterned bulk materials for gap separations $d=10,\SI{100}{nm}$, whereas HTCs from optimal bulk materials are marked in the two blue star markers, at the respective $d$. Optimal 2D plasmonic materials can even have slightly higher HTCs than those of the optimal bulk plasmonic materials, with almost exactly the same scaling with the gap distance, but with their oscillator amplitudes optimized at each specific $d$. Optimal 2D conductivities range from $n_{\rm 2D}=\SI{9e9}{cm^{-2}}$ for $d = \SI{10}{nm}$ to $n_{\rm 2D}=\SI{9e13}{cm^{-2}}$ for $d = \SI{1000}{nm}$, assuming linear Dirac electronic dispersion, as epitomized in graphene. For nonideal materials such as n-type InP and n-type Si, deep-subwavelength patterning helps them approach the optimal HTC values at the cost of having to deal with tiny feature sizes. Furthermore, the accuracy of the EMT calculations can be questionable below $d= 50\ \rm{nm}$ and the triangles are therefore kept hollow. Permittivity and conductivity parameters of real materials are from references~\cite{LD_BB_metal,AZO_GZO_extrapolate,AlN_phonon,IR_Si_2009,si_doping,III-V_FIR,cdody_2015,thin_film_2009_sic,sio2_constants,optical_graphene}.}
  \label{fig:d_HTC}
\end{figure*}

For a 2D material, causality implies a conductivity of the form (cf. {\SM}),
\begin{equation}
\sigtwod = i\epsn\omega t\sum_{i=1}^N \frac{\wBi^2}{\omega^2-\wi^2+i\omega\gamma_i} , \\
    \label{eq:2d_sigma}
\end{equation}	
which is the analog of the bulk-material representation, \eqref{eps_repre}. The $t$ is a dummy-variable thickness (assumed to be $\SI{1}{nm}$ throughout) such that the $\wBi$ have dimensions of frequency, and is canceled in the sum rule for $\wBi$: $\sum_i \wBi^2 = \frac{1}{t}\frac{n_{\rm{2D}}e^2}{\epsb\meff}$~(\citeasnoun{christensen_2019}). 2D materials, including 2D insulators (e.g. undoped 2D hBN~\cite{tunable_2D_BN}), 2D semiconductors (e.g. 2D transition metal dichalcogenides like $\rm{MoS_2}$~\cite{optical_MoS2}, and 2D black phosphorous~\cite{optical_black_phosphorous}), 2D semimetals (e.g. graphene~\cite{plasmon_graphene}, borophene~\cite{borophene_conductivity}) and 2D metals (e.g. atomically thin metals~\cite{tunable_2D_metal,ultrathin_Ag}), exhibit 2D phonons/interband transitions~\cite{Narang_2d_phonon}, leading to the Drude--Lorentz terms (with non-zero $\wi$), and the latter two classes, in addition, also possess 2D plasmons/intraband transitions, leading to Drude terms (with $\wi=0$)~\cite{2D_nanophotonics_review,graphene_FIR_UV}. By the same process as for bulk materials, we use gradient descent to optimize over the set of conductivities represented by \eqref{2d_sigma}, and we find analogous results: the optimal profile is dominated by a single-pole conductivity, as shown in \figref{optimization_2D}(a), and compared to the initial guess, the optimal profile provides spectral HTC contributions at a rather broad range of low-energy frequencies, as shown in \figref{optimization_2D}(b). Although the dominant oscillator is not strictly Drude type, its frequency is nearly zero ($\approx \SI{10}{meV}$). The other oscillators either have very small amplitudes $\wBi$ if their oscillator frequencies are small, or have exactly $\wBi=0$ if their oscillator frequencies are large. The optimal HTC level is $\SI{2.7e5}{W/(m^2 K)}$, larger even than the optimal bulk-material HTC. Even for a completely 2D plasmonic material, i.e., with single 2D Drude oscillator optimized for $d=\SI{10}{nm}$ so that $\omega_{\rm{b,0}}=\SI{0.24}{eV}$, HTC can be as large as $\SI{2.4e5}{W/(m^2 K)}$, as highlighted by the blue marker in \figref{optimization_2D}(c). For every other $d$, the optimal Drude oscillator amplitudes $\omega_{\rm b}$, and hence 2D carrier concentrations $n_{\rm 2D}$, need to be re-optimized in order to match the optimal gap surface wave modal dispersions, as opposed to being essentially constant in the case of bulk Drude materials. Yet interestingly, the optimized 2D material parameters provide HTCs that scale with $d$ exactly the same way as the optimal bulk material, as seen from \figref{d_HTC}. We detail the gap-distance dependence of the optimal $\omega_{\rm b}$ and $n_{\rm 2D}$ of 2D plasmonic materials in the {\SM}. These findings suggest that these structures may be approaching some universal, material-dimension-independent fundamental limit.

The bulk-material intuition of \secref{lineshape} carries over to 2D materials: the Drude pole, or Drude--Lorentz pole with negligible $\wn$ ($\ll \SI{0.1}{eV}$) is ideal because it maximizes the bandwidth of large RHT response, and the optimal loss rate is a moderate value that trades off strong confinement (low loss) with large bandwidth (high loss). In addition, 2D material performances are also affected by the substrate. Substrates with refractive index higher than that of vacuum lead to decreased confinement (cf. {\SM}) and are therefore nonideal. The Wien frequencies for RHT and HTC are quite similar to those of bulk materials, with a slight blue shift due to the wider bandwidth arising from the dispersion of two sheets of 2D materials at the optimal loss rate, as shown in \figref{overlap}(a,c). We include a plot of the optimal 2D-material Wien frequencies in the {\SM}.


\section{Candidate materials for Maximum near-field RHT}
\label{sec:realmats}
Here we synthesize the optimal bulk- and 2D-material results of the previous sections to identify the best candidate materials for maximal HTC and RHT. For bulk-material HTC at \SI{300}{K}, we found three key properties: Drude-like response with resonant wavelengths of $\approx\SI{19}{\mu m}$, small background permittivity and moderate loss rates ($g \approx$ 0.02 to 0.18, decreasing as $\epsb$ increases). Loss rates and background permittivities can be tabulated for a wide variety of materials. In \figref{epsb_wr} we plot HTC level curves as a function of resonant frequency $\wr$ and background permittivity $\epsilon_{\rm b}$, overlaid with a wide variety of possible materials. Noble metals (Ag, Al) and alternative plasmonic materials such as GZO (Ga-doped ZnO) and Dy-doped CdO have resonant frequencies that are too large, while semiconductors such as Si and GaAs can be doped to the right resonant frequencies but exhibit background permittivities that are too large. Lightly doped TCOs and III-Nitrides are particularly promising material classes that present themselves as potentially optimal.

\begin{figure*}[htb]
  \includegraphics[width=0.8\linewidth]{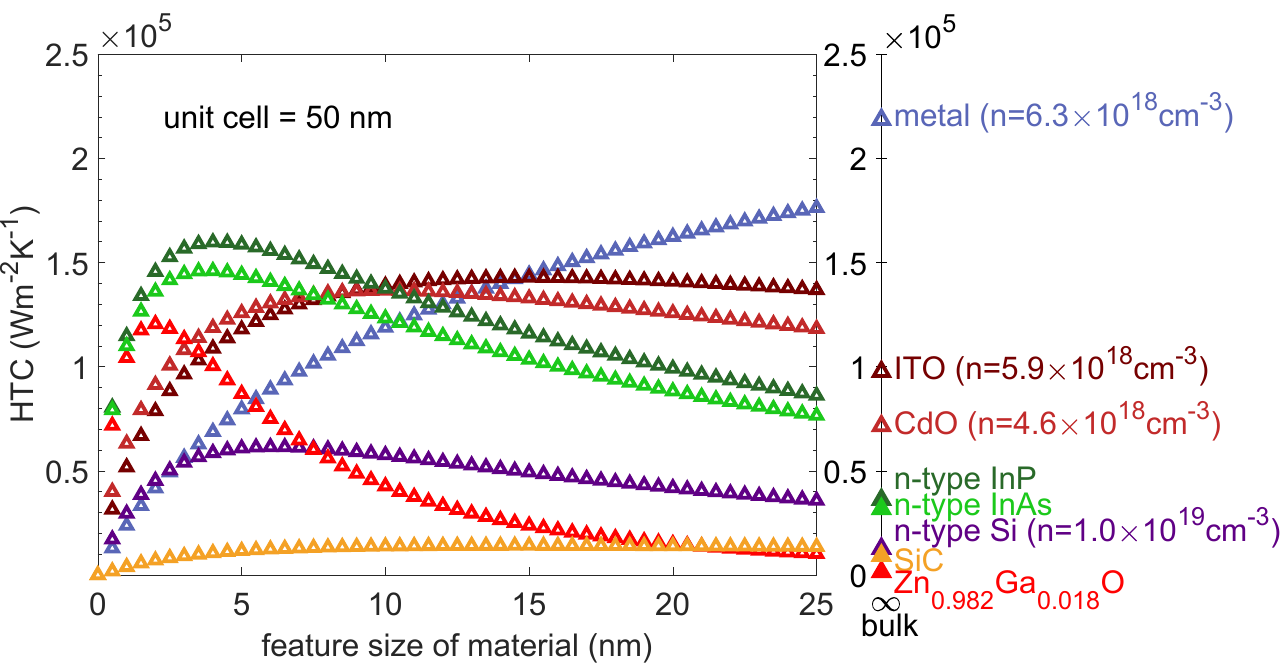}
  \caption{HTCs from periodic cylindrical-hole patterning of many materials, as a function a feature size for a unit cell fixed at \SI{50}{nm} length and width. For ideal materials (metal, ITO, CdO at their optimal carrier concentrations), the unpatterned HTC (right-most axis) reaches at least half of the best patterned HTC levels. For materials with much higher than ideal carrier concentrations (n-type InP, n-type InAs, n-type Si, $\rm{Zn_{0.982}Ga_{0.018}O}$), patterning can greatly help. Patterning does not help polar dielectrics (SiC). Solid triangles indicate materials with experimentally measured permittivity data at the required carrier concentrations, while hollow triangles indicate the need for such measurements. All finite-feature-size data uses effective-medium theory and may overestimate HTC, further strengthening the case for ideal materials without patterning.}
  \label{fig:feature}
\end{figure*}

\Figref{d_HTC} compares the theoretical HTC values for many bulk and 2D materials, as well as the optimal possible values, which are shown as the dashed (bulk) and dotted (2D material) lines. The previous theoretical state-of-the-art for unpatterned bulk materials, using $\SiO$, is depicted with orange markers. Doped III-V's such as GaAs and InP, even with their high background permittivities, can already show enhancements beyond $\SiO$, given published permittivity data ~\cite{III-V_FIR,sio2_constants}. Yet the real gains to be had are with the lighter materials, such as AlN, CdO, and ITO. If these materials can be doped to the optimal carrier concentrations listed in \figref{d_HTC}, they can exhibit 5X enhancements beyond the current state-of-the-art. (The optimal carrier concentrations tend to range from 3 to \SI{7e18}{cm^{-3}} multiplied by $\epsb \meff / m_e$.) Patterning in sub-optimal materials such as n-type InP and n-type Si can lead to slight further enhancements, discussed below, though the feasibility of effective-medium theory for describing such response is dubious at separations below \SI{50}{nm} for the given choice of temperature and unit cell size, as indicated by the open markers. 2D materials are particularly promising: purely plasmonic response can exhibit broad resonant bandwidth and strong confinement at the same time. For example, graphene with Fermi level $\EF=\SI{0.4}{eV}$ (optical conductivity from\citeasnoun{optical_graphene}) is shown to exhibit near-optimal HTC values at separations on the order of \SI{100}{nm}, though a key distinction from the optimal bulk materials is that the optimal oscillator amplitude $\omega_{\rm b}$ and therefore the optimal 2D carrier concentration $n_{\rm 2D}$ vary with the separation distance. Agnostic of the bandgap and electronic dispersion of the 2D materials, the optimal $\omega_{\rm b}$ and the corresponding maximal HTCs can be identified for each $d$ as shown in \figref{optimization_2D}(c) and {\SM}. These HTCs along the dotted grey line in \figref{d_HTC} are higher than those from any other materials. For linear Dirac electronic dispersion, as in graphene, the optimal carrier concentration optimal $n_{\rm 2D}$ is $\SI{9e11}{cm^{-2}}\times \left({\frac{d}{\SI{100}{nm}}}\right)^2$. Graphene holds potential for optical response over a great range of electromagnetic spectrum from radio waves to visible frequencies~\cite{graphene_FIR_UV}, and its optical conductivity $\sigma_{\rm 2D}(\omega)$ is well-studied in literature, with the infrared spectral range dominated by intraband transitions~\cite{optical_graphene,plasmon_graphene,measurement_graphene}, making it the exemplary 2D material in our study. However, as the optical properties of other 2D plasmonic materials, 2D semiconductors and semimetals in particular, are better characterized, it is likely that they may offer similar or superior performance to graphene.

\Figref{feature} confirms the pros and cons of metamaterial patterning. In \secref{patterning} we found that periodic cylindrical holes offer strong performance in a lithography-compatible form factor, and in \figref{feature} we consider the effects of such patterns on HTC values at \SI{300}{K}. For a unit cell size of \SI{50}{nm}, about the largest possible that could conceivably exhibit effective-medium behavior at the separations of interest in the near field, we find that HTC for each material tends to peak at feature sizes on the order of \SI{5}{nm}. For these feature sizes the significant air fraction reduces the background permittivity and increases the HTC bandwidth. Yet one can see that these effects are marginal, and that the optimal bulk values (hollow triangles on the right-hand axis) offer nearly the same HTC values in much simpler architectures.

\section{Looking forward}\label{discussion}
In this article, we identified the optimal material characteristics for maximum near-field RHT rate and HTC. Our results suggest two key avenues for future exploration: synthesis and characterization of mid- to far-infrared Drude plasmonic materials with small background permittivities, and identification of optimal patterning schemes outside of the realm of effective-medium theory.

The first avenue, centering around the development of mid- to far-infrared plasmonic materials, stems from the three properties that we identify as critical to maximal HTC for bulk materials: small background permittivity (\secref{bgepsilon}), moderate loss rate (\secref{loss}), and a resonance frequency (where $\Re \epsilon \approx -1$) corresponding to $\approx \SI{19}{\mu m}$ wavelength at \SI{300}{K}, and which scales linearly with temperature (\secref{wien}). TCOs and III-Nitrides with moderate carrier concentrations should be nearly ideal, and validation will be important as there has been little investigation into engineering plasmonic material properties at such long wavelengths. Moreover, since moderate loss is superior to low loss, new materials should be available that might traditionally have been too lossy for other plasmonics applications.

Meanwhile, 2D plasmonic materials with a single-Drude-pole optical conductivity (and thus negligible high-frequency Drude--Lorentz poles) can offer record-level near-field HTCs at their optimal carrier concentrations and loss rates. 2D semiconductors and 2D semimetals with strong 2D plasmons and the ability to support highly-confined broadband resonances in the infrared may be great materials to start with. Our analysis suggests engineering efforts devoted to optimizing the spectral bandwidth and broadband confinement of gap surface resonances of 2D materials near the prescribed near-field Wien frequencies, through either doping, gate-biasing, introducing heterostructures and nano-patterning. Being naturally surface passivated and thus easy for various integration methods with existing building blocks and devices~\cite{vanderwaals_integration,integration_2D}, 2D materials offer great promise in future energy technologies. Serendipitously, these bulk and 2D material candidates we propose not only offer possibilities to maximize near-field RHT efficiencies but also provide platforms for tunable thermal applications with the array of available switching and dynamic control approaches~\cite{dynamic_nfrht,gate_tunable_nfrht,graphene_switching,2D_modulator,ito_modulator,Vasudev13_integrate_enz,integrate_gan,iii_n_led}.

The second avenue to explore is that of wavelength-scale patterning, which lies between the two regimes studied in this paper (bulk materials and sub-wavelength-scale patterning of metamaterials). We showed that subwavelength-scale patterning, resulting in effective-medium properties, can lead to further gains in maximal RHT and HTC, but that such gains come at the expense of very small features relative to their bulk counterparts and only very modest rate increases. Thus for wavelength-scale patterning the open questions are two-fold: will the larger size scale of the patterning relieve the stringent feature-size constraints imposed by effective-medium theory, and, in tandem, will it enable substantially larger HTCs and RHT rates beyond the values predicted here? Given the significant recent work in both large-scale, computational ``inverse design'' ~\cite{Miller2012,inversedesign_review,yao2020_antenna,Chung20_highNA,jin2020inverse}, as well as analytical and computational bounds to optical response~\cite{Miller2015,Miller2016material,Shim2019bandwidth,kuang2020maximal,limited_role_structuring,comp_bounds}, conclusively answering such questions should be feasible in the near future.

\section{Acknowledgments}
This work was supported by the Army Research Office under grant number W911NF-19-1-0279.

\bibliography{materials_NFRHT}

\end{document}


\title{Supplementary Material: Optimal materials for maximum near-field radiative heat transfer}

\author{Lang Zhang}
\affiliation{Department of Applied Physics and Energy Sciences Institute, Yale University, New Haven, Connecticut 06511, USA}
\author{Owen D. Miller}
\affiliation{Department of Applied Physics and Energy Sciences Institute, Yale University, New Haven, Connecticut 06511, USA}

\date{\today}

\maketitle

\tableofcontents

\section{Permittivity and conductivity representations via discretized Kramers--Kronig relations}
In this section we show that starting only with the Kramers--Kronig (KK) relations (in turn, dependent only on causality and basic asymptotic-frequency assumptions), a general representation of \emph{any} permittivity is that of a summation of Drude--Lorentz oscillators. We start with the KK relation, one version of which relates the real part of the permittivity at any one frequency to a principal-value integral of the imaginary part:
\begin{equation}
\Re\left(\epsilon(\omega)\right)=1+\frac{2}{\pi}{\rm P.V.}\int_0^{\infty}\frac{\omega'  \Im \epsilon(\omega^{\prime})\rm{d}\omega^{\prime}}{\omega^{\prime 2}-\omega^2}.
\label{eq:KK}
\end{equation}
One can see that the integrand already has a passing resemblance to a lossless Lorentz--Drude oscillator with oscillator frequency $\omega'$, but it may not be immediately obvious how to think about the $\omega' \Im \epsilon(\omega')$ term in the numerator. The key is to use a known sum rule for the imaginary part, which is~\cite{sum_rule_chi}:
\begin{equation}
    \int_0^{\infty}\omega^{\prime}\Im\left(\epsilon(\omega^{\prime})\right)\,{\rm d}\omega' = \frac{\pi\wp^2}{2},
    \label{eq:sum_rule}
\end{equation}
where $\wp$ is the plasma frequency, which is related to free carrier concentration through Eq. (3) in the main text. The sum rule implies that $\omega' \Im \epsilon(\omega')$ can be discretized into local basis functions whose amplitudes are constrained by the constant on the right-hand side of \eqref{sum_rule}. As one example, we consider the limit of perfectly localized delta-function basis functions for the decomposition:
\begin{align}
\omega^{\prime}\Im\left(\epsilon(\omega^{\prime})\right)=\frac{\pi \wp^2 }{2}\sum_i^N c_i \delta(\omega^{\prime}-\omega_i).
\label{eq:expansion_integrand}
\end{align}
The constant $\pi \omega_p^2 / 2$, in tandem with the sum rule of \eqref{sum_rule}, ensures that basis-function coefficients that determine the distribution of $\omega' \Im \epsilon(\omega')$ must sum to 1:
\begin{align}
    \sum_i^N c_i = 1.
\end{align}
Inserting this representation of the imaginary part into the KK relation for the real part, \eqref{KK}, we find:
\begin{equation}
    \Re\left(\epsilon^{\rm KK}(\omega)\right)=1+\sum_i^N \frac{c_i \wp^2}{\omega_i^2-\omega^2},
    \label{eq:KKreal}
\end{equation}
where the superscript ``KK'' has been added to distinguish from a Drude--Lorentz representation below. Hence in the limit of perfectly localized basis functions, the KK relations imply a permittivity whose real part satisfies \eqref{KKreal}, and whose imaginary part (rewriting \eqref{expansion_integrand}) satisfies
\begin{equation}
    \Im\left(\epsilon^{\rm KK}(\omega)\right)=\frac{\pi\wp^2}{2}\sum_i^N c_i \frac{\delta(\omega-\omega_i)}{\omega_i}.
    \label{eq:KKimag}
\end{equation}

It is straightforward to show that the KK representation of \eqreftwo{KKreal}{KKimag} is equivalent to a lossless Drude--Lorentz (DL) representation. Consider a permittivity comprising a summation of Drude--Lorentz oscillators:
\begin{align}
    \epsilon^{\rm DL}(\omega) = \left(1+\sum_i^N\frac{c_i\wp^2}{\omega_i^2-\omega^2-i\gamma\omega}\right).
\end{align}
In the limit of lossless oscillators, we take $\gamma \to 0$ from above. To evaluate the DL expression in this limit, we use the identify $\lim_{\gamma\to 0} \left[1/(x+i\gamma)\right] = 1/x - i\pi \delta(x)$, in which case the DL expression simplifies to:
To evaluate the limit, considering the real-line integral of $\omega_i$ gives an important identity:
\begin{equation}
    \lim_{\gamma\to 0}\frac{1}{\omega_i^2-\omega^2-i\gamma\omega}=\frac{1}{\omega_i^2-\omega^2}+i\pi\delta(\omega_i^2-\omega^2) = \frac{1}{\omega_i^2 - \omega^2} + \frac{i\pi}{2\omega_i} \delta(\omega_i - \omega),
    \label{eq:lim0}
\end{equation}
where the last equality uses the fact that $\omega_i \geq 0$ to remove an additional delta-function term proportional to $\delta(\omega + \omega_i)$. Inserting the right-hand side of \eqref{lim0} into the DL representation gives:
\begin{equation}
    \lim_{\gamma\to 0}\epsilon^{\rm DL}(\omega)=1+\sum_i^N\frac{c_i\wp^2}{\omega_i^2-\omega^2}+i\frac{\pi\wp^2}{2}\sum_i^N  c_i \frac{\delta(\omega-\omega_i)}{\omega_i}  = \epsilon^{\rm KK}(\omega).
\end{equation}
Hence, any KK material can be represented as a (possibly infinite) sum of Drude--Lorentz oscillators.

A similar KK representation and sum rule~\cite{sum_rule_graphene} yields an analogous Drude--Lorentz representation for arbitrary 2D conductivities (with dummy thickness variable $t$ inserted for dimensionality compatibility):
\begin{align}
    \sigma(\omega)=i\epsn\omega t \left[ \sum_i^N\frac{c_i\wp^2}{\omega_i^2-\omega^2}+i\frac{\pi}{2}\sum_i^N \wp^2 \frac{c_i}{\omega_i} \delta(\omega-\omega_i)\right] = \lim_{\gamma\to 0}\sigma^{\rm DL}(\omega).
\end{align}

\section{Setup of oscillator-strength optimization of HTC}
Given the general permittivity/conductivity expressions derived in the previous section, arising from the Kramers--Kronig relations, the optimal material that maximizes HTC is the one whose oscillator strengths satisfy the optimization problem:
\begin{equation}
\begin{aligned}
& \underset{\wAi,\wi}{\text{maximize}}
& & \text{HTC} \\
& \text{subject to}
& & \wAi\geq0, \wi \geq 0.
\end{aligned}
\end{equation}
To reduce the dimensionality of the problem, we can select a finite number of oscillators $N$ and fix their intrinsic frequencies $\wi$ over a one-dimensional grid, leaving only the oscillator strengths $\wAi$ to be optimized. (By fixing $\omega_{0,0} = 0$ we ensure the first oscillator is the Drude type, and any other $\omega_{0,i\neq0}$ represents the oscillator frequencies of Drude-Lorentz oscillators.) We can then optimize over the $\wi$ via gradient descent, converging on local optima and re-running the optimizations many times to ensure discovery of solutions at or very near the global optima.

To compute the gradients, we start with the modal photon exchange function:
\begin{align}
    {\xi(\omega,\beta)} = \frac{4(\Im r )^2e^{2i\knz d}}{|1-r^2e^{2i\knz d}|^2},
\end{align}
and the interface reflectivity:
\begin{align}
    r(\omega,\beta)=\frac{\knz \epsilon-\kz}{\knz \epsilon+\kz},
\end{align}
where $\knz=\sqrt{\frac{\omega^2}{c^2}-\beta^2}$ and $\kz=\sqrt{\epsilon\frac{\omega^2}{c^2}-\beta^2}$ are respectively the z-component of wavevectors in air and in material. The gradient of the modal exchange function is given by
\begin{align}
    \frac{{\rm d} \xi(\omega,\beta)}{{\rm d} \wAi} = 2 \Re \left[\frac{{\rm d} \xi(\omega,\beta)}{{\rm d} r} \frac{{\rm d} r(\omega,\beta)}{{\rm d} \epsilon} \frac{{\rm d} \epsilon(\omega)}{{\rm d} \wAi} \right],
\end{align}
where
\begin{align}
    & \frac{{\rm d} \xi(\omega,\beta)}{{\rm d} r}=\frac{\phase(r-r^*)(1-rr^*p)}{2(1-r^*p)(1-r^2p)^2},
    \\
    & \frac{{\rm d} r(\omega,\beta)}{{\rm d} \epsilon}=\frac{2\kz\knz}{(\knz\epsilon+\kz)^2},
    \\
    & \frac{{\rm d} \epsilon(\omega)}{{\rm d} \wAi}=-\frac{2\wAi}{\omega^2-\wi^2+i\omega\gamma_i}.
\end{align}
The HTC gradients with respect to each oscillator strength are calculated by numerical integration:
\begin{align}
    \frac{{\rm dHTC}}{{\rm d} \wAi}=\frac{1}{4\pi^2}\int_0^{\infty}\dtheta \int_0^{\infty}\frac{{\rm d} \xi(\omega,\beta)}{{\rm d} \wAi}\beta{\rm d}\beta {\rm d}\omega.
\end{align}

\section{Results from oscillator-strength optimization}
\begin{figure}[htb]
  \includegraphics[width=0.81\textwidth]{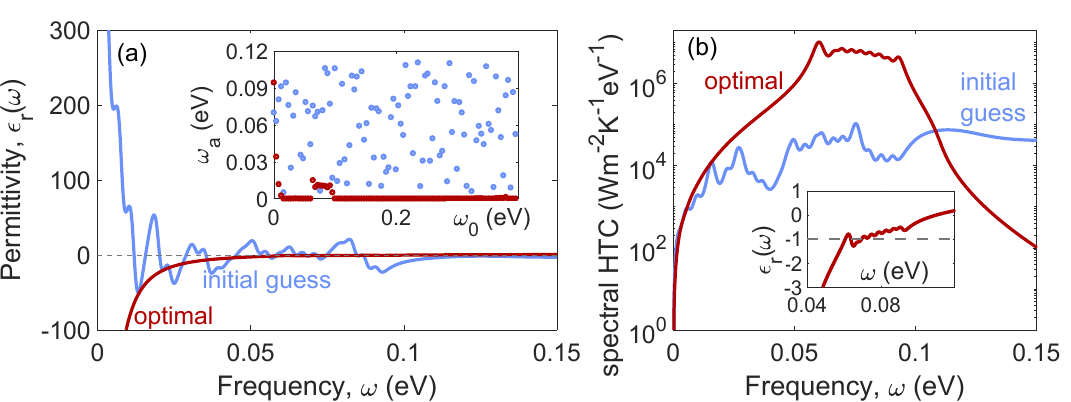}
  \caption{(a) The initial guess (blue) and optimal permittivity profile (red) of one representative 100-oscillator optimization, with oscillator frequency $\wi$ sampled from $0$ to slightly beyond $\SI{0.4}{eV}$. Inset shows the corresponding oscillator amplitudes $\wAi$. (b) Spectral HTC from the initial (blue) and optimal (red) permittivities. Inset: zoomed-in profile of the optimal $\epsilon_{\rm r}$ near the optimal $\wr$.}
  \label{fig:optimization_N100}
\end{figure}

In the main text, we presented the results from a 400-oscillator optimization where grid spacing of $\wi$ is \SI{0.001}{eV} and $\gamma=\SI{0.004}{eV}$, which clearly shows that (1) the Drude-dominant lineshape is pivotal, and that (2) multiple Drude-Lorentz oscillators can be useful to broaden the resonance bandwidth by flattening the range of $\epsilon \approx -1$ near the optimal $\wr$.

At lower resolutions and the same loss levels, it becomes obvious that the individual oscillators with oscillator frequency $\wi$ near the optimal $\wr$ tend to create small bumps in the permittivity profile and in spectral HTC, as can be demonstrated in a 100-oscillator optimization where $\wi$ is sampled from $0$ to $\SI{0.4}{eV}$ with grid spacing \SI{0.004}{eV}. From a random starting point where all $\wAi\neq 0$, the optimization result has the largest non-zero amplitude for the Drude oscillator and only a few small amplitudes for Drude-Lorentz oscillators at low frequencies. Such features from low-resolution results are retained in higher-resolution optimizations, and can be useful to help the latter better converge to high-quality local optima via successive refinement. The inset in \figref{optimization_N100}(a) shows that every $\wAi$ with oscillator frequency $\wi>\SI{0.1}{eV}$ tends to vanish. Hence the resultant lineshape is predominantly that of a single-pole Drude permittivity with $\omega_{\rm{a},0}=\wp=\SI{0.094}{eV}$ superimposed by small bumps near $\wr$ which flatten the range of $\epsilon \approx -1$ and thus broaden the resonance bandwidth. Furthermore, the bumps around the optimal $\wr$, due to the coarse sampling of $\wi$ and the small loss level, can be smoothed out at higher resolution, as is apparent from the optimization results in the main text.

\section{Near-field Wien frequencies}
In this section, we derive the optimal plasma frequency, or related resonance frequency, of the Drude materials that are optimal for HTC. Using the optimal permittivity lineshape argued in the main text, i.e. Drude-type with $\epsb=1$, the only free parameter in fixing $\epsilon(\omega)$ is the plasma frequency $\wp$, or the resonance frequency $\wr$. This optimal resonance frequency problem is quite similar to the Wien frequency argument in blackbody radiation. For largest spectral contribution to HTC between blackbodies, the Wien frequency $\wienHTC$ maximizes $\dtheta B(\omega,T)$ where the spectral radiance $B(\omega,T)$ is given by
\begin{align}
    B(\omega,T)=\frac{\omega^2}{4\pi^3c^2}\frac{\hbar\omega}{e^{\hbar\omega/\kb T}-1}.
    \label{eq:Planck}
\end{align}
Setting $x = \frac{\hbar\omega}{\kb T}$ and taking its derivative to $0$, the maximum is found to solve the equation
\begin{align}
    (x-4)e^x+x+4=0,
\end{align}
which has the positive solution $3.83$. For Wien frequency for RHT $\wienRHT$, which maximizes $\left(\Theta(\omega, T_2)-\Theta(\omega, T_1)\right) B(\omega,{\rm T})$, and taking the cold-body, or environment temperature $T_1=0$, the corresponding equation is
\begin{align}
    (x-3)e^x+3=0,
\end{align}
which has the positive solution $2.82$. Take the radiation from the sun for example: with $T_2=\SI{5778}{K}$, one finds $\wienRHT=\SI{1.91}{eV}$, which is the well-known Wien frequency from solar radiation. Parametrizing Planck's law in \eqref{Planck} by wavelength $\lambda$ gives a different Wien frequency $\wienRHT=\SI{1.40}{eV}$.

For near-field Wien frequencies, the major contribution comes from gap surface waves (tunnelling evanescent waves) instead of propagating plane waves. This is apparent in the much sharper spectral photon exchange of the ``near field, Drude" configuration than the ``far field, blackbody" configuration, and lead to Wien frequencies significantly smaller than those of the blackbodies. By normalizing the frequencies in the HTC integral against the resonance frequency $\nu = \omega / \wr$, the HTC is given by
\begin{align}
        {\rm HTC} = \frac{1}{4\pi^2}\int_0^{\infty}\frac{{\rm d}\Theta(\nu,T)}{{\rm d}T}\wr \underbrace{\int_0^{\infty}\xi(\nu,\beta)\beta{\rm d}\beta}_{S(\nu)} {\rm d}\nu.
\end{align}
$S(\nu)$ is now independent of $\wr$ and sharply peaks exactly at $\nu=1$, i.e. $\omega=\wr$. We therefore arrive at a single-parameter optimization aiming to maximize $\frac{{\rm d}\Theta(\nu=1,T)}{{\rm d}T}\wr$ over $\wr$. The optimization simplifies to solving
\begin{align}
    (x-3)e^x+x+3=0,
    \label{eq:wien_bulk}
\end{align}
leading to $x=\frac{\hbar\omega}{\kb T}=2.57$.
As for RHT in the near field, with the cold-body temperature $T_1=0$, the objective function now is $\left(\Theta(\nu=1, T_2)-\Theta(\nu=1, T_1=0)\right) \wr=\Theta(\nu=1, T_2) \wr$, leading to
\begin{align}
    (x-2)e^x+2=0,
\end{align}
which has a solution $x=\frac{\hbar\omega}{\kb T_2}=1.59$. For other non-zero $T_1$, similar equations can be derived and solved, and the results are plotted in Fig.6(b) of the main texts.

Just like the far-field case for blackbodies, the near-field Wien laws conclude that the hotter the temperature, the bluer the peak spectral thermal radiation, owing to the Planck distribution $\ttheta$. The linear scaling relation with temperature is retained in the near field, which is a signature of thermal exchange via radiation, in contrast to conduction, whose scaling relation we show in the next section. The differences between the density of states of propagating plane waves in the far field and the local density of states of the gap surface waves in the near field manifest themselves as the different numerical factors in these transcendental equations, which result in much smaller solutions as Wien frequencies in the near field. Fortuitously, the equations for the near-field case are agnostic whether parametrizing the RHT or HTC integral with $\omega$ or $\lambda$, preventing the ambiguity of the spectral location of Wien peak which occurs in the far-field case.

\section{Comparison between optimal radiative HTC and conductive HTC}
Given the optimal loss rate for a given separation distance, $\wienHTC$ derived above dictate the optimal $\wp$ at every temperature, and lead to optimal near-field radiative HTC values that scale linearly with temperature as described by Eq. (9) in the main text, which we include here:
\begin{equation}
    {\rm HTC_{\rm rad}}= \left(\SI{760}{W/m^2/K^2}\right) T.
\end{equation}
For conductive HTC for the plane-plane configuration, gap separations on the order of 10nm are well below the mean free path in air $\SI{68}{nm}$~\cite{mean_free_path}, thus it is safe to assume ballistic conductive heat transfer: no collision of gas molecules happen in the air gap, and the only collisions occur on the air-material interface. The conductive HTC for such configuration is~\cite{many_body_rht}:
\begin{equation}
    {\rm HTC_{\rm cond}}=\frac{3}{2}\kb n_{\rm g}v_{\rm z},
\end{equation}
where $v_{\rm z}=\sqrt{\frac{\kb T}{m}}$ is the mean velocity of particles normal to the plates, and $m$ is single-particle mass. Assuming particle number density in air under standard conditions for temperature and pressure, $n_{\rm g}=\SI{2.5e25}{m^{-3}}$, the conductive HTC scales with the square root of $T$ as
\begin{align}
    {\rm HTC_{\rm cond}} = \left(\SI{8786}{W/m^2/K^{3/2}}\right) \sqrt{T}.
\end{align}
At lower temperatures, conductive heat transfer dominates, whereas near-field RHT can surpass conductive heat transfer rates at high enough temperatures. For the gap separation of $\SI{10}{nm}$, the two HTCs cross at $T=\SI{134}{K}$.

\section{HTC from three types of effective-medium metamaterials}

\begin{figure}[htb]
  \includegraphics[width=0.81\textwidth]{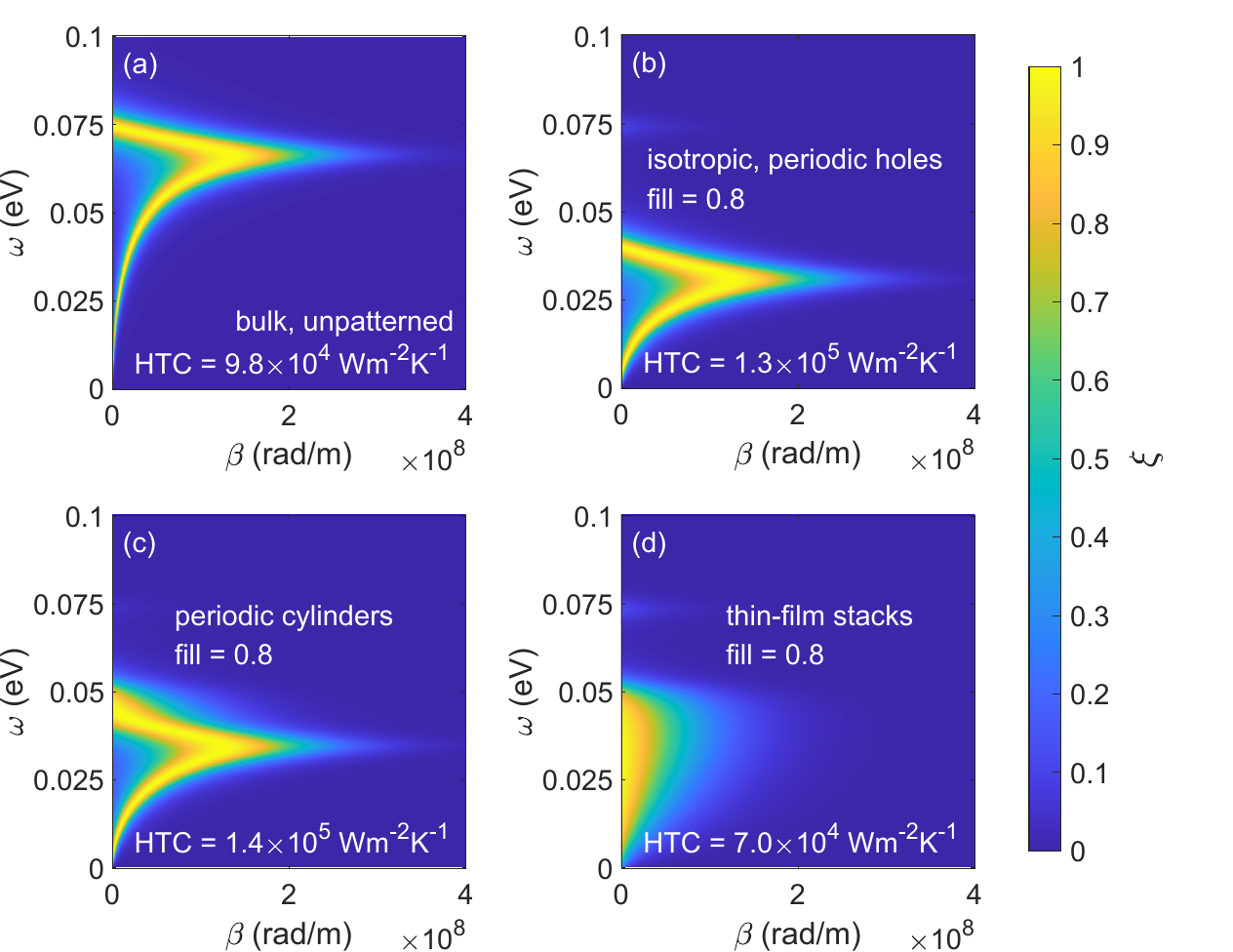}
  \caption{Modal photon exchanges of ITO (a) in bulk, unpatterned form, (b) with periodic and deep-subwavelength patterning of isotropic holes, (c) with periodic and deep-subwavelength patterning of cylinders in z-direction, and (d) with periodic and deep-subwavelength patterning of thin-film stacks in z-direction (hyperbolic metamaterials). The volume fill ratio of air is 0.8 for all the patterned materials.}
  \label{fig:modal_ITO}
\end{figure}

Many insights about the effective-medium patterned materials can be gained by examining in tendem the homogenized material parameters and the corresponding modal photon exchanges. Here we choose ITO ($\epsb = 3.8$) at its optimal carrier concentration ($n = \SI{5.9e18}{cm^{-3}}$) as an example. As can be seen from \figref{modal_ITO}(a), for the bulk, unpatterned ITO, the optimal $n$ guarantees alignment of strong gap surface resonance at the optimal $\wr$, namely $\wwien$. But because of the non-unity $\epsb$, its relative bandwidth, $\frac{\Delta\omega}{\wr}$ ($\Delta\omega$ here is the full width at half of the maximal spectral HTC), is not as broad as it would optimally be based only on causality. Broader bandwidth can be expected from inclusions of air holes (or air cylinders, air layers) which reduces the effective $\epsb$ by reducing the effective electron density.

According to Maxwell--Garnett effective-medium theory, an air fill fraction $f$ in a medium with permittivity $\epsilon$ yields an effective permittivity of
\begin{equation}
\epsilon_{\rm{eff}} = \frac{(1-f)\epsilon+1+f}{1-f+(1+f)\epsilon}\epsilon. \\
    \label{eq:nanohole}
\end{equation}
From \figref{modal_ITO}(b), the surface resonance frequency is lowered compared to the unpatterned case in (a), due to smaller $\omega_{\rm p, eff}$, and the relative bandwidth is broadened, due to smaller $\epsilon_{\rm b, eff}$. Although $\wr$ becomes misaligned with $\wwien$ after patterning, the overall HTC is higher than that of the unpatterned ITO, thanks to the reduction in $\epsilon_{\rm b, eff}$.

Periodic patterning of cylinders along z-direction, the in-plane patterning version of holes, has effective permittivity along the ordinary axis the same as \eqref{nanohole}, and along the extraordinary axis:
\begin{equation}
\epseeff = (1-f)\epsilon+f.
    \label{eq:nanohole_e}
\end{equation}
Anisotropy leads to hyperbolic resonances near $0.4 \sim \SI{0.5}{eV}$, as shown by \figref{modal_ITO}(c), yet other than that, much of the behavior is similar to that from isotropic, periodic hole patterning. This modal photon exchange corresponds to the data point of ITO at feature size $\SI{25}{nm}$ in Fig. 10 of the main text. The enhancement of HTC here, about a factor of 1.4, is close to that from the optimal feature size.

Hyperbolic metamaterials, in this case thin stacks of material and air arranged periodically in z-direction, can be homogeneously characterized by
\begin{align}
&\epsoeff = (1-f)\epsilon+f,
    \label{eq:nanolayer_o} \\
&\epseeff = \frac{\epsilon}{1-f+f\epsilon},
    \label{eq:nanolayer_e}
\end{align}
along ordinary and extraordinary axes respectively. For such configuration, thermal exchanges occur predominantly though hyperbolic resonances, shown by \figref{modal_ITO}(d), whose confinement to the planar-planar surfaces is fairly small. Thus even with a broad resonance bandwidth, its HTC is nowhere near that from the above two types of effective-medium metamaterials, and even smaller than that of the unpatterned ITO.

We chose ITO (at its optimal free-carrier concentration) as the material here due to its near-optimality for near-field RHT. However, this material is not necessarily optimized for a particular subwavelength patterning strategy. In fact, as shown in Fig. 10 in the main text, for deep-subwavelength patterning of periodic cylinders at $\SI{4}{nm}$ feature size, n-type InP and InAs can outperform the optimal ITO. However, even the HTC from the best effective-medium metamaterial (that from n-type InP in our investigation) is less than a factor of 2 improvement from the bulk, unpatterned optimal value provided by ITO, and smaller than the that from bulk ideal metal. Clearly, under effective material parameters, whether isotropic or anisotropic, nearly the best HTC values can be provided by simple bulk materials.

Moreover, the predicted HTCs from effective-medium approximations can only hold as gap distances are adequately large, temperatures adequately low, and unit-cell sizes and feature sizes adequately small. They are likely overestimates when the gap distances are small.

\section{2D plasmonic materials: reflectivities, dispersion relations, resonant frequencies and near-field HTCs}
At the air interface of a 2D polaritonic medium with conductivity $\sigma(\omega,\beta)$ with substrate $\epssub$, the reflectivity is given by
\begin{equation}
    r(\omega,\beta)=-1+\frac{\ksubz}{\knz}\frac{2 \epsn \omega}{2\epssub \epsn \omega+\ksubz \sigma(\omega,\beta)},
\end{equation}
while the 2D surface wave dispersion is given by~\cite{plasmon_graphene}
\begin{equation}
    \frac{\epssub}{\ksubz}+\frac{1}{\knz}=-\frac{\sigma(\omega,\beta)}{\omega \epsn},
    \label{eq:2D_ingle_dispersion}
\end{equation}
where $\ksubz=\sqrt{\frac{\epssub \omega^2}{c^2}-\beta^2}$.

Following the notations from the main text, a simple 2D plasmonic conductivity with negligible spatial dispersion can be written as
\begin{equation}
    \sigma(\omega)=i\epsn\omega t\frac{\wb^2}{\omega^2+i\gamma\omega}.
    \label{eq:sigma_wb}
\end{equation}
The 2D surface wave dispersion in the non-retarded regime is
\begin{equation}
    \beta=\frac{2}{t}\frac{\omega^2+i\gamma\omega}{\wb^2},
\end{equation}
which shows the scaling of $\wr$ with $\sqrt{\beta}$. More precisely, through a more rigorous quasi-static analysis, it can be established that~\cite{christensen_2019}
\begin{equation}
    \wr^2=\frac{\beta t}{2}\wb^2.
\end{equation}

For two parallel sheets of identical 2D material in the near-field regime with separation $d$, air as substrate, the gap surface waves have dispersion
\begin{equation}
    \left(\frac{2}{\kn}+\frac{i\sigma}{\omega\epsn}\right)^2e^{i\knz d}-\left(\frac{i\sigma}{\omega\epsn}\right)^2e^{-i\knz d}=0.
\end{equation}
With the simple 2D conductivity model as \eqref{sigma_wb} and the optimal parameters, these 2D gap surface waves provide modal photon exchange as shown in \figref{modal_2D}(a), which also leads to the 2D material spectral photon exchange profile in Fig. 5 of the main text.

We also compute the HTC from $\sigtwod$ of 2D materials reported in literature, such as graphene whose intraband electronic transitions dominate its optical properties in the infrared:
\begin{equation}
    \sigma^{\rm intra}(\omega,T)=\frac{2ie^2T}{\pi\hbar(\omega+i\gamma)}{\rm ln}\left(2{\rm cosh}\frac{\EF}{2T}\right),
    \label{eq:sigma_intra}
\end{equation}
where $\EF$ is the Fermi level, and $\EF=\SI{0.4}{eV}$ gives to the graphene HTC in Fig. 10 of the main text.

\section{Wien frequencies for 2D plasmonic materials}
\begin{figure}[htb]
  \includegraphics[width=0.4\textwidth]{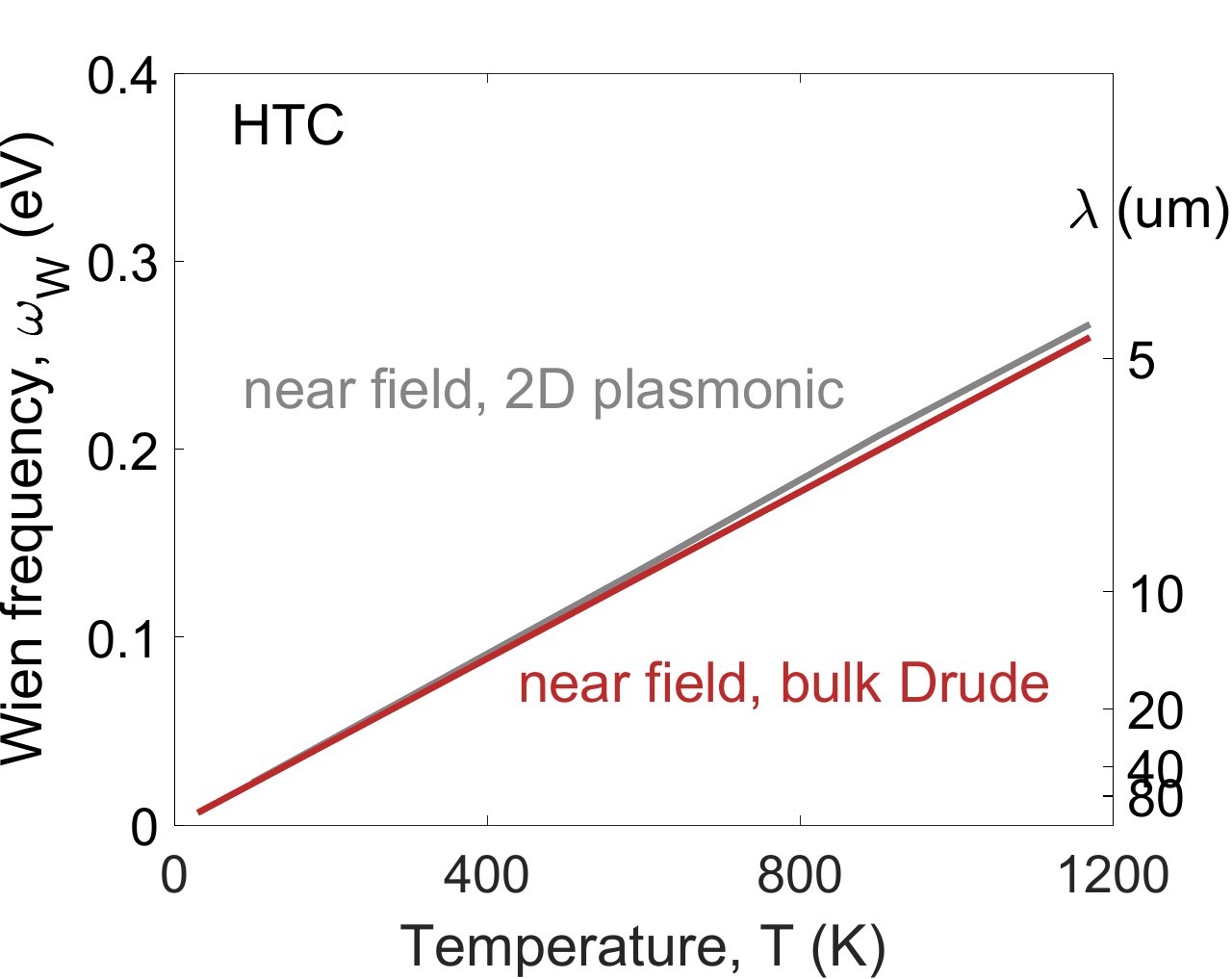}
  \caption{Wien frequencies of 2D and bulk plasmonic materials for the near-field planar-planar configuration. The red curve for the ``bulk Drude'' case can be analytically derived (\eqref{wien_bulk}), whereas the grey curve for ``2D plasmonic'' is numerically identified for each $T$.}
  \label{fig:wien_2D}
\end{figure}

Although the 2D plasmons can have quite different modal dispersions from the bulk ones, they are by nature still narrow-band compared to the dispersion and density of states for propagating plane waves. Therefore, to a large degree, the Wien analysis similar to that of the optimal bulk Drude materials can apply. To be exact, we numerically identified $\wwien$ for 2D plasmonic materials, and plot them in tandem with those of the bulk Drude ones in \figref{wien_2D}. At their optimal loss rate which is around $g=0.06$ for the temperature range of interest, the HTC Wien frequencies of 2D materials are quite similar to those of bulk Drude materials, with a slight blue shift due to the slightly wider bandwidth of the dispersion from two sheets of 2D materials.

\section{2D plasmonic materials with different substrate refractive indices}
\begin{figure}[htb]
  \includegraphics[width=0.75\textwidth]{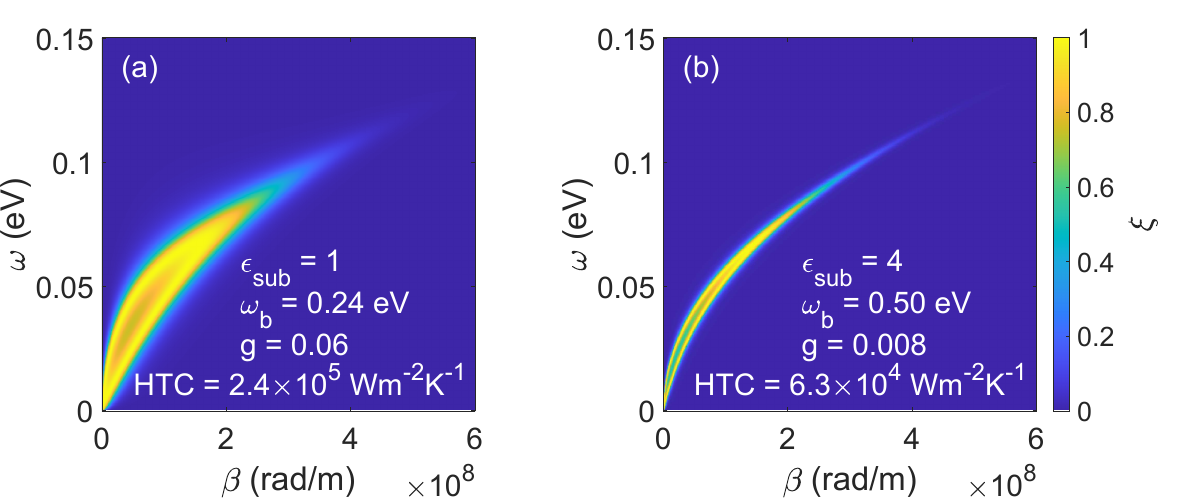}
  \caption{Modal photon exchange of the optimal 2D plasmonic material at $d=\SI{10}{nm}$, for (a) air substrate and (b) substrate with $\epssub=4$.}
  \label{fig:modal_2D}
\end{figure}

The 2D surface wave dispersion and therefore RHT can be quite sensitive to the substrate permittivity for 2D materials. Without much loss of generality, we assume that the substrate has negligible polaritonic response in the frequency range we are interested in, and thus can be modeled by a constant permittivity, or refractive index.
As a general rule of thumb, the higher the substrate refractive index, the higher the optimal $\wb$ and the lower the optimal $g$. As can be seen from the comparison between \figref{modal_2D}(a) and (b), where both are using optimal parameters for their respective $\epssub$, a larger $\epssub$ leads to much narrower modal bandwidth of 2D surface waves, thus smaller spectral contribution to HTC at the resonant frequencies. More generally, more resonances and better confinement can be supported had there been larger index contrast, which makes vacuum the best substrate in theory for the 2D materials of interest here.

\section{Gap distance dependence of optimal $\wb$ and $n_{\rm 2D}$ for 2D plasmonic materials}
\begin{figure}[htb]
  \includegraphics[width=0.5\textwidth]{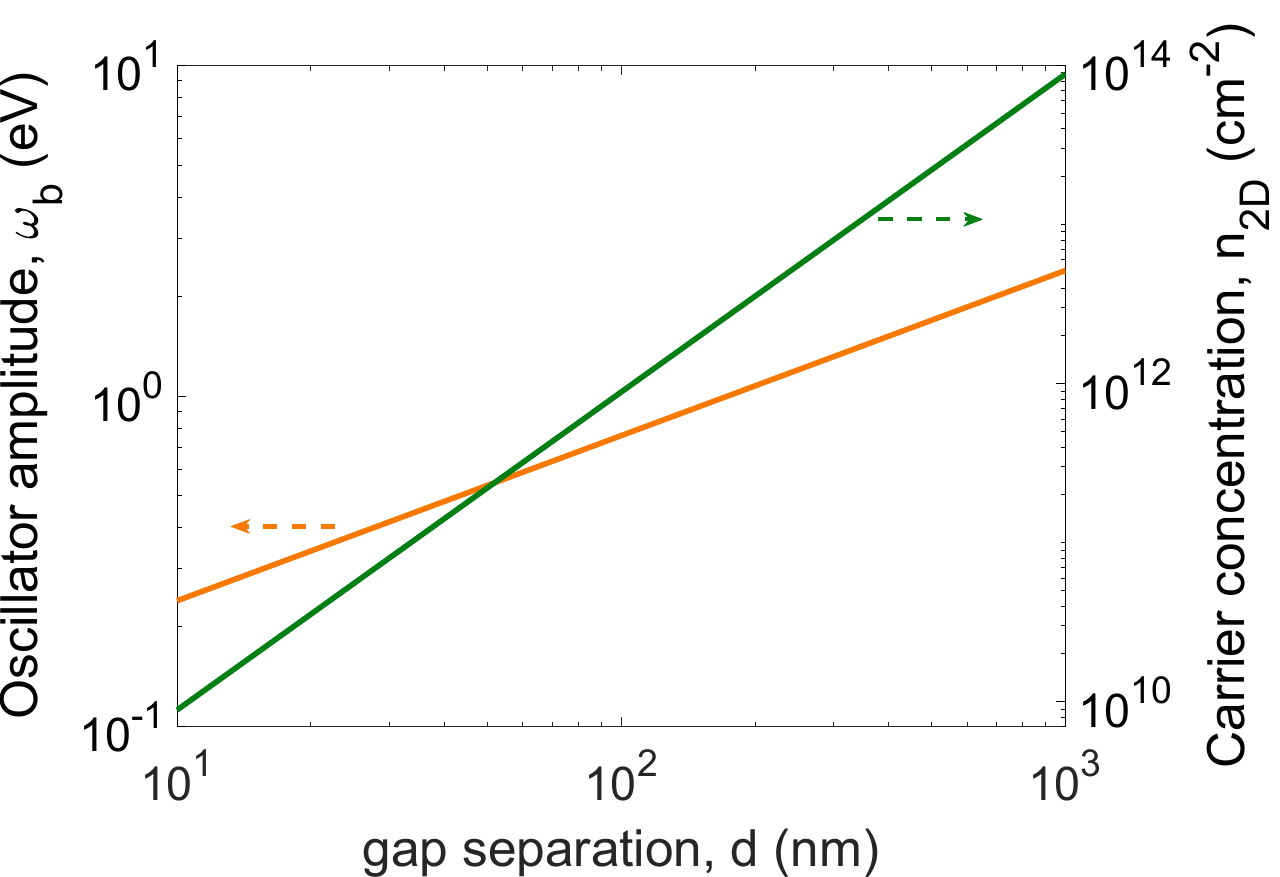}
  \caption{Optimal $\wb$ (assuming dummy-variable thickness $t=\SI{1}{nm}$) and $n_{\rm 2D}$ (assuming Dirac linear dispersion) for 2D plasmonic materials at different gap separations for temperature $T=\SI{300}{K}$ and substrate $\epssub=1$.}
  \label{fig:distance_2D}
\end{figure}

We parametrically optimize $\wb$ and $g$ pairs for the 2D conductivity in \eqref{sigma_wb} for every gap separation $d$, and find the ones maximizing HTC. In \figref{distance_2D} we present these optimal parameters for $d$ ranging from 10 to \SI{1000}{nm} that lead to the optimal ``2D plasmonic" curves in the main text. Although the scaling factor depends on the dummy-variable thickness $t$, which is assumed to be $\SI{1}{nm}$ throughout, the scaling relation of optimal $\wb \sim \sqrt{d}$ is general no matter what the electronic dispersion. As for the carrier concentration for materials with Dirac linear dispersion, as we shown in \figref{distance_2D}, optimal $n_{\rm 2D} = \SI{9e11}{cm^{-2}}\times \left({\frac{d}{\SI{100}{nm}}}\right)^2$, independent of the assumed length scale $t$.

\bibliography{materials_NFRHT}